\documentclass[aps,prb,twocolumn,amsmath,amssymb,nofootinbib,superscriptaddress,floatfix]{revtex4-1}

\usepackage{ulem}
\usepackage{amsmath,amsfonts,amssymb}
\usepackage{amssymb}
\usepackage{amsthm}
\usepackage[dvipdf]{color}
\usepackage{graphicx}
\usepackage[title]{appendix}
\usepackage{dcolumn}
\usepackage{bm} 
\usepackage[rightcaption]{sidecap}

\begin{document}
\title{Fully-Polarized Topological Isostatic Metamaterials in Three Dimensions}

\author{Zheng Tang}
\thanks{These authors contribute equally to this work.}
\affiliation{Key Lab of Advanced Optoelectronic Quantum Architecture and Measurement (MOE), School of Physics, Beijing Institute of Technology, Beijing, 100081, China}

\author{Fangyuan Ma}
\thanks{These authors contribute equally to this work.}
\affiliation{Key Lab of Advanced Optoelectronic Quantum Architecture and Measurement (MOE), School of Physics, Beijing Institute of Technology, Beijing, 100081, China}

\author{Feng Li}
\email{phlifeng@bit.edu.cn}
\affiliation{Key Lab of Advanced Optoelectronic Quantum Architecture and Measurement (MOE), School of Physics, Beijing Institute of Technology, Beijing, 100081, China}

\author{Yugui Yao}
\affiliation{Key Lab of Advanced Optoelectronic Quantum Architecture and Measurement (MOE), School of Physics, Beijing Institute of Technology, Beijing, 100081, China}

\author{Di Zhou}
\email{dizhou@bit.edu.cn}
\affiliation{Key Lab of Advanced Optoelectronic Quantum Architecture and Measurement (MOE), School of Physics, Beijing Institute of Technology, Beijing, 100081, China}

\begin{abstract}
Topological surface states are unique to topological materials and are immune to disturbances. In isostatic lattices, mechanical topological floppy modes exhibit softness depending on the polarization relative to the terminating surface. However, in three dimensions, the polarization of topological floppy modes is disrupted by the ubiquitous mechanical Weyl lines. Here, we demonstrate, both theoretically and experimentally, the fully-polarized topological mechanical phases free of Weyl lines. Floppy modes emerge exclusively on a particular surface of the three-dimensional isostatic structure, leading to the strongly asymmetric stiffness between opposing boundaries. Additionally, uniform soft strains can reversibly shift the lattice configuration to Weyl phases, reducing the stiffness contrast to a trivially comparable level. Our work demonstrates the fully-polarized topological mechanical phases in three dimensions, and paves the way towards engineering soft and adaptive metamaterials. 
\end{abstract}

\maketitle

\emph{Introduction---}Isostatic structures~\cite{fruchart2020dualities, PhysRevLett.129.204302}, also known as Maxwell structures~\cite{PhysRevLett.110.198002, kane2014topological}, are mechanical frames that perfectly balance the degrees of freedom and constraints. These structures, ranging from molecular to architectural-scales~\cite{phillips1979topology}, are viewed as networks of nodes and links. Their significance lies in providing insights into stability and adaptability for innovative material and structure design, particularly in soft-matter systems~\cite{lubensky2015RRP, PhysRevLett.124.225501, PhysRevLett.129.125501, huang2023jammed, bertoldi2017NRM}. Isostatic structures host zero-frequency edge modes that exhibit topological protection~\cite{rocklin2017NC, wang2023JMPS, coulais2017N, ma2023PRL}, because the boundary mechanical softness and rigidity remain unchanged even when disturbances and damage occur~\cite{xiu2023PNAS, paulose2015NP, chen2014nonlinear}.

In one- and two-dimensional isostatic lattices, mechanical floppy modes can be ``topologically fully-polarized", as they emerge exclusively on a single boundary, while the opposing surface is completely devoid of floppy modes. This behavior results in a highly asymmetric contrast of boundary stiffness in a uniform structure. Fully-polarized isostatic lattices establish the connection between elasticity and topological electronic band theory~\cite{RevModPhys.82.1959, PhysRevLett.95.146802, PhysRevLett.42.1698, PhysRevLett.61.2015, RevModPhys.83.1057, PhysRevLett.98.106803}, laying the foundation for topological mechanics~\cite{susstrunk2015observation, vakakis2001normal, ruzzene2019PRL, nash2015topological, ruzzene2019PRB, rosa2023NJP, tempelman2021PRB, bertoldi2015PRL, paulose2015PNAS, Qiu2023PRL, tempelman2024modal, souslov2017NP}. However, for two-dimensional isostatic lattices, this conceptual correspondence is not applicable to \emph{out-of-plane} motions, which lack topological protection and mechanical polarization.

In three-dimensional (3D) isostatic lattices, the fully-polarized topological mechanics is disrupted by the ubiquitous Weyl lines~\cite{Olaf2016PRL, Huber2017AM} that close the mechanical bandgap and reduce the contrast of boundary stiffness to a trivially comparable level. Recent studies have theoretically proposed 3D isostatic lattices that eliminate Weyl lines~\cite{Olaf2016PRL, Vitelli2017PNAS}, but these designs allow floppy modes to emerge on both opposing surfaces, resulting in a stiffness contrast that is still trivially comparable. Furthermore, in the precedented experiments~\cite{bergne2022EML, Huber2017AM}, the continuous mechanical junctions introduce finite bending stiffness that pushes the 3D-printed specimens beyond the isostatic point~\cite{mao2020PRL, PhysRevLett.122.248002}, making topological numbers undefined. Thus, the elimination of Weyl lines and full polarization of mechanical topology in 3D isostatic lattices remains challenging.

{In this work, we demonstrate, both theoretically and experimentally, the fully-polarized topological mechanical phase in 3D. Using the three-dimensional example known as the generalized pyrochlore lattice, we illustrate this topological mechanical phase and the resulting distinctive boundary elasticity.} Our analytic design principle is based on the mechanical transfer matrix~\cite{Zhou2018PRL, zhou2019PRX} that polarizes all floppy modes to concentrate on a single open boundary of the lattice, whereas the opposite surface is clear of floppy modes. Consequently, the lattice is topologically fully-polarized and exhibits highly contrasting boundary mechanics.

Moreover, {isostatic} lattices grant uniform soft strains of the entire structure, known as Guest-Hutchinson modes~\cite{Guest2003JMPS}, that reversibly shear the lattice configuration and induce transitions among {topologically polarized and mechanical Weyl} phases. {This uniform shearing in a mechanical lattice is a nonlinear mechanism that alters the geometric configuration of all unit cells, but without inducing elastic energy.} By shearing the lattice to the mechanical Weyl phase, the contrast in local stiffness is reduced to a comparable level, {because floppy modes arise on both parallel open surfaces of the lattice. This fully-polarized mechanical phase in 3D, together with the freely switchable topological transition, results in advancements not possible in 2D systems~\cite{pyrochloreSM}, such as topological softness bounded to 3D dislocations, static mechanical non-reciprocity in all spatial dimensions, and topologically protected all terrain tire. }

\begin{figure}[htbp]
\centering
\includegraphics[scale=0.46]{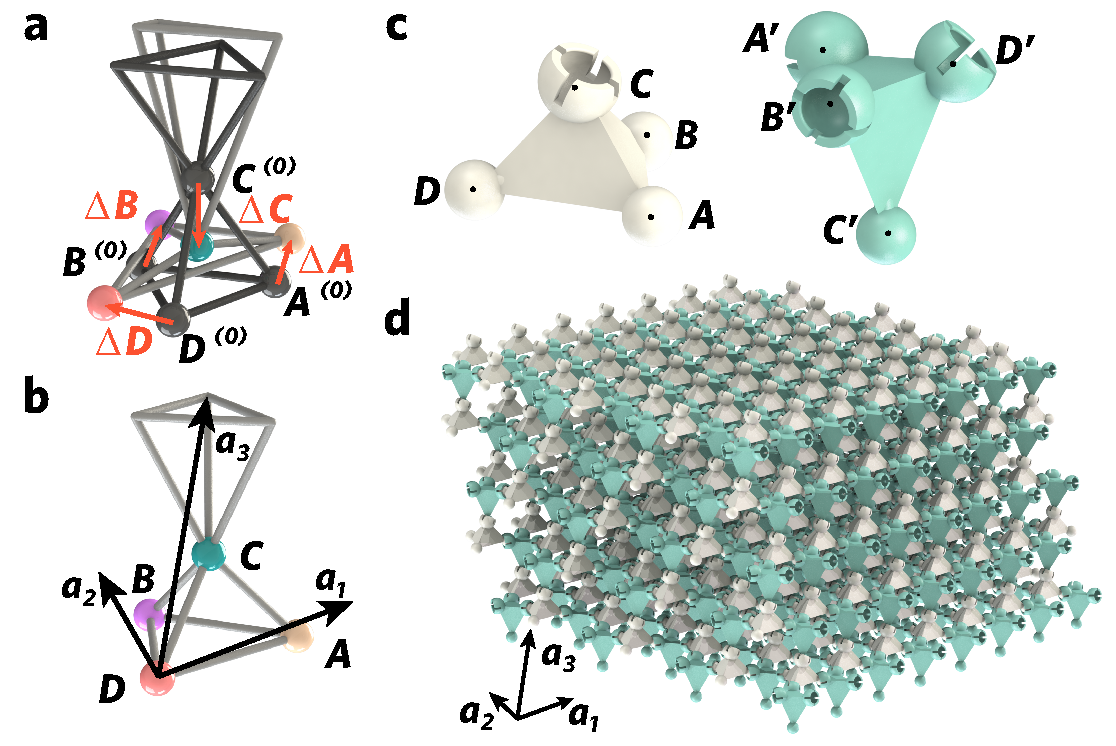}
\caption{Design principle of the 3D fully-polarized topological mechanical metamaterial. \textbf{a} and \textbf{b} describe the unit cells of the regular and generalized pyrochlore lattices, respectively, with the same primitive vectors $\bm{a}_1$, $\bm{a}_2$, and $\bm{a}_3$. Each cell contains 12 bonds and 4 vertices marked by $\bm{A}$, $\bm{B}$, $\bm{C}$, $\bm{D}$. \textbf{c}, the white and blue rigid bodies stand for the bottom and top tetrahedra that form the unit cell of the mechanical metamaterial. Each vertex is equipped with either a concave or convex surface, constituting the spherical hinges that allow for relative free rotations. \textbf{d}, the assembly of the fully-polarized topological mechanical metamaterial.
}\label{fig1}
\end{figure}

\emph{3D fully-polarized topological {isostatic} lattices---}Our prototype is a polymer 3D mechanical structure that uses pyrochlore lattice for network connectivity. Fig. \ref{fig1} displays two geometries of pyrochlore lattices, namely the regular and generalized ones in \textbf{a} and \textbf{b}, respectively. Fig. \ref{fig1}\textbf{a} shows the unit cell of the regular pyrochlore lattice, with four corners at $\bm{A}^{(0)} = \ell(1,1,0)/2$, $\bm{B}^{(0)} = \ell(0,1,1)/2$, $\bm{C}^{(0)} = \ell(1,0,1)/2$, and $\bm{D}^{(0)} = \ell(0,0,0)/2$; primitive vectors $\bm{a}^{(0)}_1 = \ell(1,1,0)$, $\bm{a}^{(0)}_2 = \ell(0,1,1)$, and $\bm{a}^{(0)}_3 = \ell(1,0,1)$; and a length scale of $\ell=24$ mm. In Fig. \ref{fig1}\textbf{b}, the geometry of the generalized pyrochlore lattice deviates from the regular one, with vertex positions $\bm{X} = \bm{X}^{(0)}+\Delta \bm{X}$ for $\bm{X} = \bm{A}, \bm{B}, \bm{C}, \bm{D}$ and primitive vectors $\bm{a}_i = \bm{a}^{(0)}_i+\Delta \bm{a}_i$ for $i=1,2,3$. These geometric parameters {reside in a vast 14-dimensional space} (demonstrated in SI~\cite{pyrochloreSM}), {which poses challenges for searching fully-polarized topological mechanical phases}. {To address this,} we employ the technique known as the 3D mechanical transfer matrix~\cite{pyrochloreSM}. {By decomposing the isostatic lattice into layers of lower dimensions, we ensure that the lattice geometry associated with the transfer matrix allows for consistent growth of mechanical floppy modes from the top to the bottom layer.} {Remarkably, this approach} substantially reduces the parameter space from 14 to just 3 dimensions. {Figure 1 illustrates a geometric example that facilitates the consistent growth of floppy modes from top to bottom. In this configuration,} we have $\Delta\bm{A}= 0.053(1, 0.3, 0)$, $\Delta\bm{B} =0.053(0.55, 0, 1)$, $\Delta\bm{C} = 0.053(-1,1,-1)$, $\Delta\bm{D} = 0.053(-1.8,-1,1.2)$, and $\Delta\bm{a}_{i=1,2,3}=0$.

The unit cell of the generalized pyrochlore metamaterial consists of two polymer tetrahedra, each featuring a spherical hinge at every vertex, as depicted in Fig. \ref{fig1}\textbf{c}. In Fig. \ref{fig1}\textbf{d}, the $\bm{A}$, $\bm{B}$, $\bm{C}$, $\bm{D}$ vertices of the white tetrahedra are connected to the $\bm{A}'$, $\bm{B}'$, $\bm{C}'$, $\bm{D}'$ tips of the green tetrahedra, enabling free rotations between neighboring bodies and eliminating bending stiffness. {Within the unit cell, the site positions of the green tetrahedron~\cite{pyrochloreSM} are give by $\bm{A}'=\bm{A}-\bm{a}_1+\bm{a}_3$, $\bm{B}'=\bm{B}-\bm{a}_2+\bm{a}_3$, $\bm{C}'=\bm{C}$, and $\bm{D}'=\bm{D}+\bm{a}_3$.} As a result, each spherical hinge provides three constraints, and each tetrahedron is connected to four hinges. A total of $(3\times 4)/2=6$ constraints are imposed on a tetrahedron, balancing the six degrees of freedom of the rigid body. Consequently, the pyrochlore metamaterial is classified as an {isostatic} lattice due to the perfectly balanced degrees of freedom and constraints~\cite{kane2014topological}. This {isostatic} point {ensures the rigorous definition of topological mechanical indices, distinguishing them from the previously undefined topological numbers in super-isostatic structures~\cite{Huber2017AM, bergne2022EML}}.

Floppy modes refer to tetrahedron movements that do not deform their rigid bodies or mutual hinges, and thus do not involve elastic potential energy. Furthermore, floppy modes occur slowly, and their zero-frequency nature makes kinetic energy negligible. These properties allow us to exactly map the Newtonian statics of the metamaterial to an idealized spring-mass network, which enables the analytic study of the topological phases in static mechanical properties. 

The idealized spring-mass model is established by assigning a mass particle to each site and representing each edge with a central-force Hookean spring. In both the spring-mass system and the pyrochlore metamaterial, the static mechanics are equivalent, as each set of floppy modes corresponds to undistorted spherical hinges, edges, and tetrahedral bodies. The static mechanics of the spring-mass pyrochlore model can be described by the compatibility matrix $\textbf{C}$, which maps site displacements to spring elongations~\cite{lubensky2015RRP}. In spatially repetitive systems, the compatibility matrix can be Fourier-transformed into reciprocal space~\cite{pyrochloreSM}, $\textbf{C}(\bm{k})$, where $\bm{k}$ represents the wavevector. This compatibility matrix defines three integer-valued winding numbers, 
\begin{eqnarray}
N_i(\bm{k}) = \frac{-1}{2\pi\mathrm{i}}\oint\limits_{\bm{k}\to \bm{k}+\bm{b}_i}d\bm{k}\cdot \nabla_{\bm{k}}\ln\det\textbf{C}(\bm{k}),\,\, i=1,2,3\,\,
\end{eqnarray}
that govern the topological phase of static mechanical properties, where the integration trajectory $\bm{k}\to\bm{k}+\bm{b}_i$ follows a straight and closed loop that is parallel to the reciprocal vector $\bm{b}_i$. 

\begin{figure}[htbp]
\centering
\includegraphics[scale=0.41]{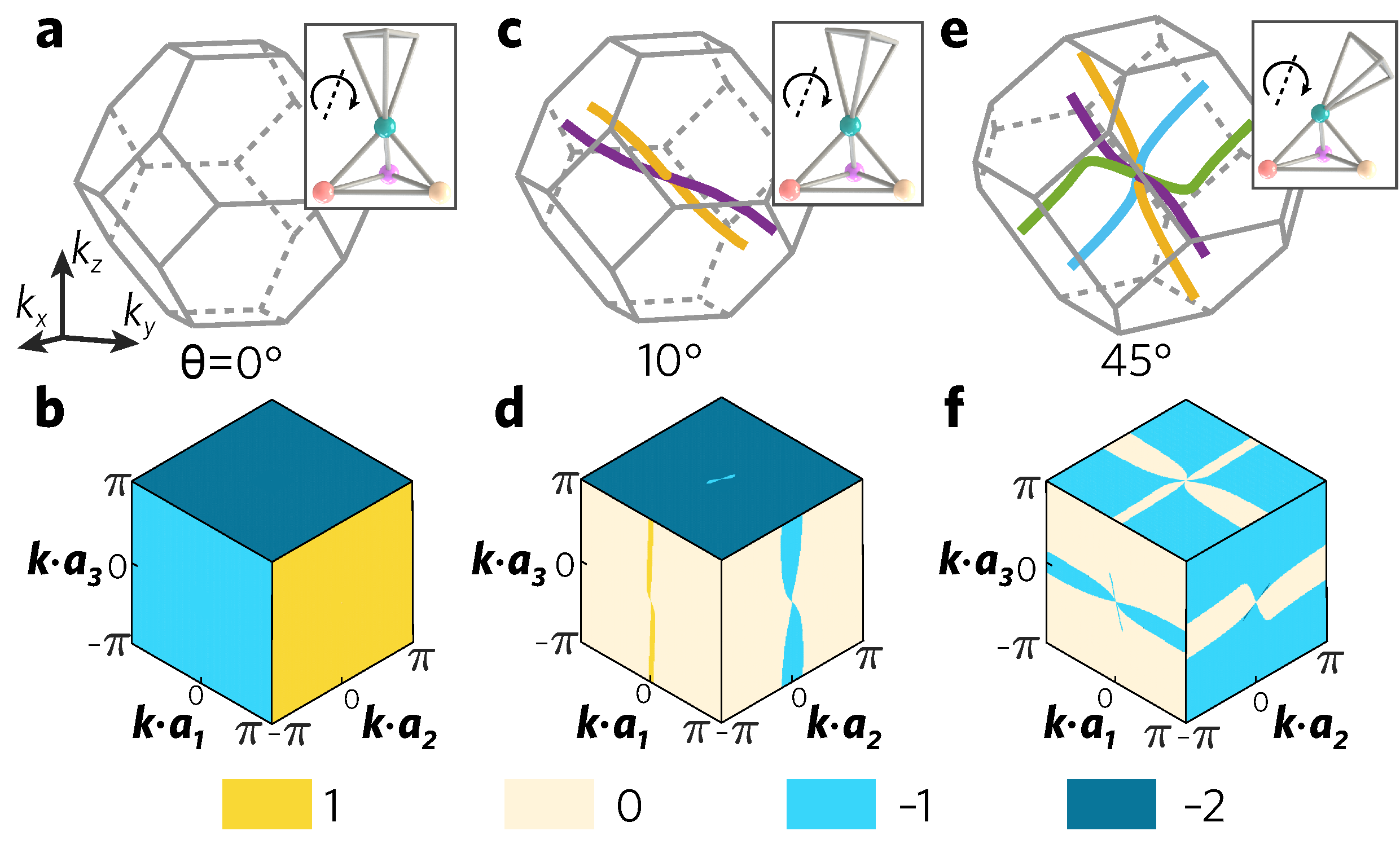}
\caption{Weyl lines and winding numbers in three configurations of the generalized pyrochlore lattice. \textbf{a}, Brillouin zone of the generalized pyrochlore lattice without mechanical Weyl lines. The unit cell of Fig. \ref{fig1}\textbf{d} is described by the inset, whose black dashed line represents the rotation axis of the uniform shearing mode. \textbf{b}, winding numbers of the surface Brillouin zones of \textbf{a}. \textbf{c} and \textbf{e} show the Brillouin zones of the generalized pyrochlore lattice for {shearing} angles $\theta = 10^\circ$ and $\theta = 45^\circ$, containing 2 and 4 Weyl lines presented by solid color curves, respectively. \textbf{d} and \textbf{f} display the winding numbers in the surface Brillouin zones corresponding to the unit cell configurations in the insets of \textbf{c} and \textbf{e}, respectively.
}\label{fig2}
\end{figure}

For different unit cell geometries, winding numbers can manifest qualitatively distinct behaviors. In most geometries of pyrochlore lattices, winding numbers exhibit jumps between integers as the wavevector moves across the Brillouin zone, and the critical boundaries between different integers are called mechanical Weyl lines. However, in models with gapped mechanical spectra, {such as the structure exhibited in Fig. \ref{fig1}}, winding numbers stay invariant for arbitrary wavevector, and the mechanical bands are clear of Weyl lines. These globally defined winding numbers constitute the vector~\cite{kane2014topological}
\begin{eqnarray}
\bm{R}_{\rm T} = \sum_{i=1}^3 N_i\bm{a}_i
\end{eqnarray}
known as the topological polarization. {This globally defined vector characterizes the topological phases of static mechanics and reflects the topological robustness of floppy modes in both the spring-mass model and the pyrochlore metamaterial.} These zero-frequency mechanisms prefer to localize on the open boundary that terminates this topological polarization, while the opposite parallel boundary has fewer floppy modes.

For the lattice configuration in Fig. \ref{fig1}, the topological polarization is $\bm{R}_{\rm T}=\bm{a}_1-\bm{a}_2-2\bm{a}_3$. {As topological polarization depends on the choice of the unit cell, we introduce the local polarization vector, denoted as $\bm{R}_{\rm L}$. This vector characterizes how nodes and bonds are locally connected on the open surface~\cite{kane2014topological, pyrochloreSM}, effectively canceling the gauge dependence of the total polarization vector, $\bm{R}_{\rm T}$. Together, these two polarizations govern the number density, $\nu$, of topological floppy modes on the open surfaces of isostatic lattices. Specifically, we have $\nu=(1/2\pi)(\bm{R}_{\rm T}+\bm{R}_{\rm L})\cdot\bm{G}$, where $\bm{G}$ represents the reciprocal vector with its normal pointing outward from the open surface. For the pyrochlore structure shown in Fig. \ref{fig1}, we find the number density of the top and bottom open surfaces as $(\nu_\uparrow, \nu_\downarrow) = (0,3)$.}

{The top boundary, with the floppy mode density $\nu_\uparrow = 0$, lacks any topological floppy modes entirely. In contrast, the bottom boundary hosts three floppy modes per supercell, as indicated by $\nu_\downarrow = 3$. Due to the contribution of these floppy modes to local softness, the top boundary remains as rigid as the lattice's interior, while the bottom surface becomes significantly softer than the interior. As the lattice thickness increases, the contrast ratio in edge stiffness grows exponentially due to the localization of topological floppy modes near boundaries. This exotic behavior arises uniquely from topological polarization in isostatic lattices, which we term ``fully-polarized topological mechanical metamaterials in 3D." In the subsequent section, we validate this highly polarized boundary elasticity both numerically and experimentally.}

It is worth emphasizing that our 3D topological lattice is fundamentally distinct from its 2D counterparts, {because 2D mechanical structures are prone to deform, distort, and lose mechanical stability due to external pressure, thermal expansion, and bending~\cite{charara2022omnimodal, zunker2021EML}. Our work discovers the fully-polarized topological mechanical phase in 3D. This new phase opens up avenues towards novel applications of topological mechanical metamaterials,} including asymmetric wave propagation, directional polar elasticity in {isostatic} media, and topological fracturing protection~\cite{pyrochloreSM}.

\emph{{Uniform Shearing} and Transformable Mechanical Weyl Phase---} {Isostatic} lattices are known to host nonlinear and uniform soft strains of the whole structure, namely Guest-Hutchinson modes~\cite{Guest2003JMPS, fruchart2020dualities, meng2020JMPS}, that reversibly shear the geometry without causing any elastic energy. In the pyrochlore metamaterial, the {uniform shearing} represents the rotation of the top tetrahedron around the spherical hinge connecting it to the bottom tetrahedron within the unit cell. This rotational mode is uniform across all unit cells in the lattice.

The insets of Figs. \ref{fig2}\textbf{c} and \ref{fig2}\textbf{e} display two configurations of the pyrochlore unit cell that demonstrate how {uniform shearing} can reversibly evolve from one state to another, as visualized by the Supplementary Video~\cite{pyrochloreSM}. {In the unit cell shown in Fig. \ref{fig2}\textbf{a}, we rotate the top tetrahedron clockwise around the spherical hinge on the axis $\bm{a}_1\times\bm{b}_3$, by angles of $10^\circ$ and $45^\circ$, to achieve the configurations depicted in the insets of Figs. \ref{fig2}\textbf{c} and \ref{fig2}\textbf{e}, respectively.} These structures display mechanical responses that are topologically distinct from those in Fig. \ref{fig2}\textbf{a}.

\begin{figure}[htbp]
\centering
\includegraphics[scale=0.41]{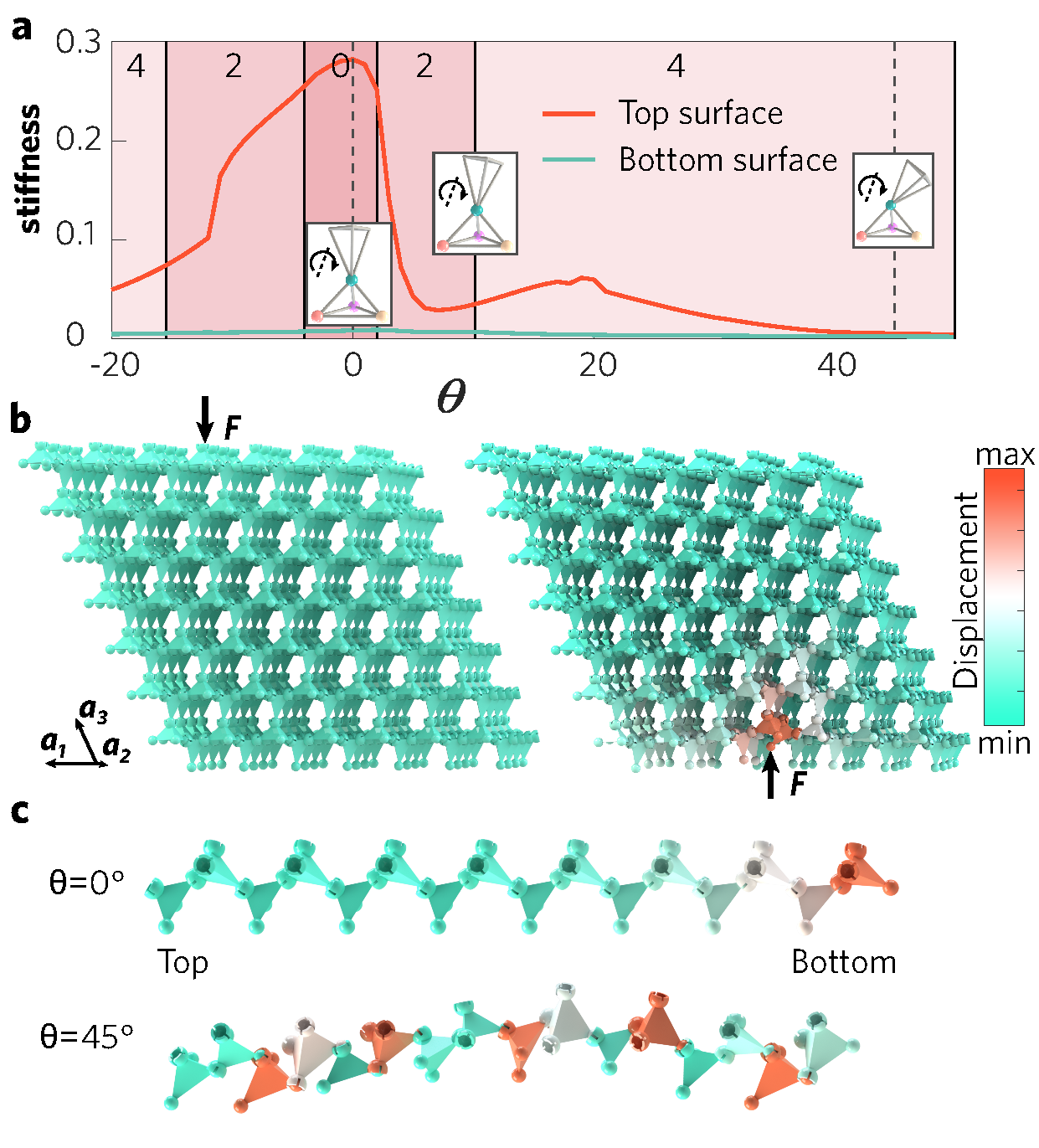}
\caption{Numerical simulations of the surface mechanics in the pyrochlore models under horizontal periodic boundaries and vertical open boundaries. \textbf{a}, the local stiffness of top and bottom open surfaces in a model composed of $5\times 5\times 8$ unit cells as the uniform shearing angle $\theta$ varies. The different background colors mark topologically distinct mechanical phases, with 0, 2, and 4 denoting topologically polarized, two-Weyl-line, and four-Weyl-line phases, respectively. \textbf{b}, top and bottom surface deformations in the left and right panels, respectively, with the colors on each tetrahedron represents its center displacement. The lattice is in the topological phase ({uniform shearing} angle $\theta=0^\circ$), and is constructed from $15\times15\times 8$ unit cells. \textbf{c}, spatial profiles of floppy modes in the topologically polarized and Weyl phases for the up and down panels, respectively. 
}\label{fig3}
\end{figure}

{We first discuss the distinct topological mechanical phases exhibited by the three unit cell configurations shown in Figs. \ref{fig2}.} In contrast to the globally-defined winding numbers in Fig. \ref{fig2}\textbf{b} that originate from the fully-polarized topological phase, the winding numbers in Figs. \ref{fig2}\textbf{d} and \ref{fig2}\textbf{f} can change as the wavevector moves~\cite{pyrochloreSM}. At the critical wavevectors where the winding numbers jump, the mechanical bandgap closes and form gapless lines in the 3D Brillouin zone. These are called mechanical Weyl lines~\cite{Vitelli2017PNAS}, and are difficult to remove due to their topological robustness. The topological charge of Weyl lines~\cite{Olaf2016PRL} is determined by a nontrivial Berry phase $N_w = \frac{-1}{2\pi\mathrm{i}}\oint_C d\bm{k}\cdot\nabla_{\bm{k}}\ln\det\textbf{C}(\bm{k})$, where the integration path $C$ encloses the gapless lines. {We highlight that the Brillouin zones depicted in Figs. \ref{fig2}\textbf{c} and \ref{fig2}\textbf{e} feature two and four Weyl lines, which correspond to a two-Weyl-line phase and a four-Weyl-line phase, respectively. 
As we discuss below, these two mechanical Weyl phases exhibit qualitatively different boundary elasticity compared to the topologically fully polarized phase.}

\begin{figure}[htbp]
\centering
\includegraphics[scale=0.41]{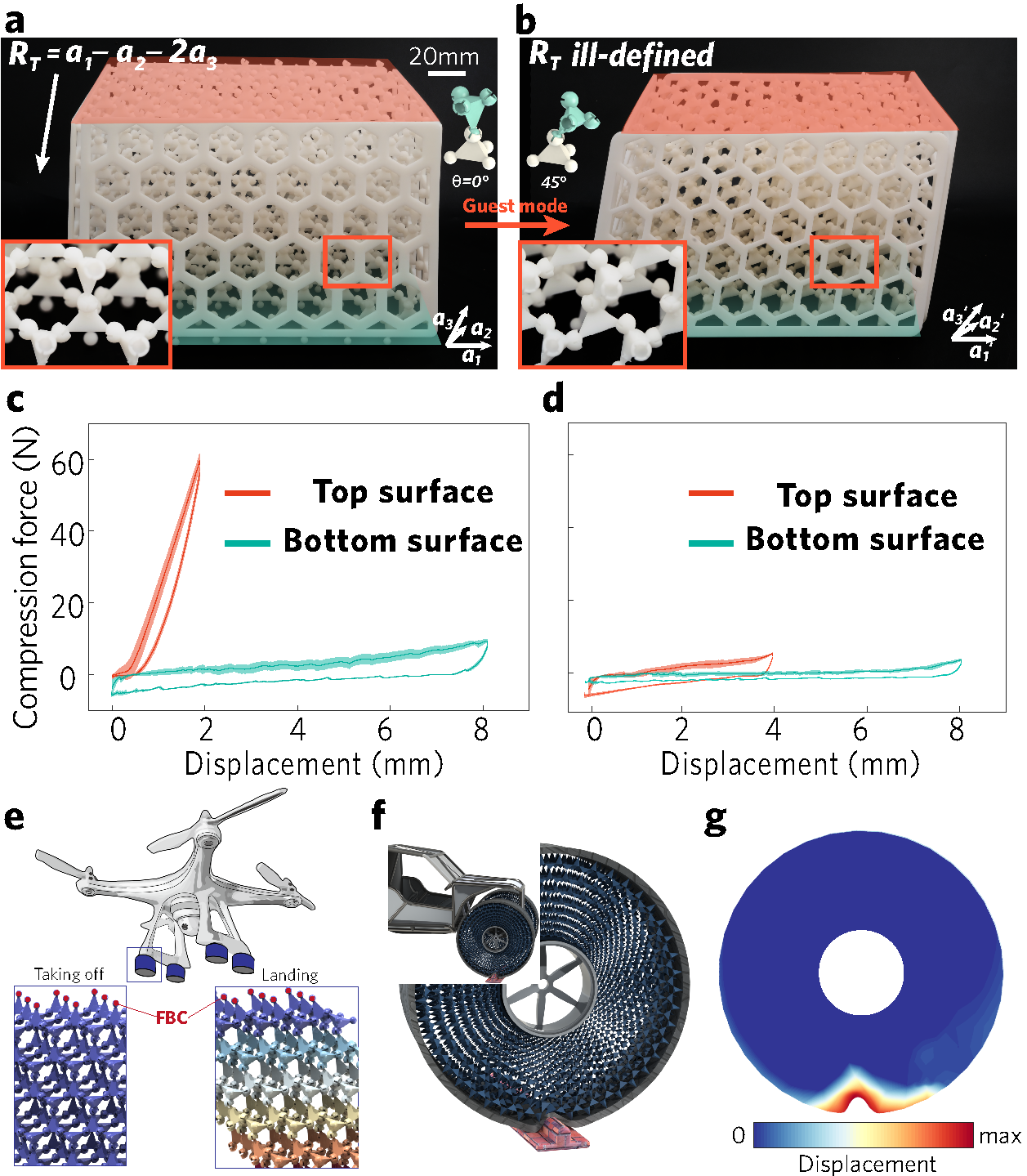}
\caption{Experimental measurements of the 3D-printed pyrochlore metamaterials. \textbf{a} and \textbf{b}, the assembly of the 3D-printed pyrochlore lattice in topologically fully-polarized and Weyl phases, respectively. The measuring open top and bottom boundaries are highlighted in red and green, whereas the four side boundaries are fixed during experimental implementations. \textbf{c} and \textbf{d}, force-displacement measurements for the topologically fully-polarized and Weyl phases, respectively, where the red and green curves represent the measurements for the top and bottom boundaries. The solid lines indicate the average values, while the shaded areas represent the standard deviation across five measurements. The positive direction of displacement is defined as normal to the measuring surface when it pushes towards the interior of the lattice. \textbf{e}, topological landing gear with reconfigurable elasticity for drones. \textbf{f}, topological pyrochlore lattice incorporated into a cylindrical domain, forming a porous wheel. \textbf{g}, numerical analysis reveals the wheel's mechanical response while rolling on a rugged terrain profile. 
}\label{fig4}
\end{figure}

{In Fig. \ref{fig3}\textbf{a}, numerical simulations are conducted to analyze the stiffness of the top and bottom open surfaces as the uniform shearing angle increased from $\theta = 0^\circ$ to $45^\circ$.} {At $\theta=0^\circ$, corresponding to the configuration in Fig. \ref{fig2}\textbf{a}, the pyrochlore metamaterial is in the topologically fully-polarized phase, exhibiting significantly higher stiffness on the top surface compared to the bottom. This strong asymmetry in surface elasticity is further corroborated by the numerical results against external poking forces shown in Fig. \ref{fig3}\textbf{b}.} {Upon increasing the uniform shearing angle to $10^\circ$, the lattice undergoes a transition into the two-Weyl-line phase, as illustrated in Fig. \ref{fig2}\textbf{c}. This transition is characterized by a significant decrease in stiffness of the top boundary, which is quantitatively supported by the numerical analysis presented in Fig. \ref{fig3}\textbf{a}. Further rotation of the uniform shearing angle to $45^\circ$ induces a shift to the four-Weyl-line phase, depicted in Fig. \ref{fig2}\textbf{e}, leading to an additional reduction in stiffness of the top open surface, as evidenced by the data in Fig. \ref{fig3}\textbf{a}. The corresponding topological mechanical phase diagram is derived in the Supplementary Information~\cite{pyrochloreSM}, exhibiting sharp and well-defined phase boundaries.}

These numerical results have a direct correspondence to the boundary experiments of the 3D-printed metamaterial. In the topologically fully-polarized case (Fig. \ref{fig4}\textbf{a}), floppy modes emerge exclusively on the bottom boundary, whereas the clearance of floppy modes on the top boundary indicates a topologically protected rigidity. This unprecedented and strongly contrasting boundary stiffness in 3D is experimentally demonstrated using force-displacement measurements in Fig. \ref{fig4}\textbf{c}. {We rotate the lattice’s unit cell configurations by a uniform shearing angle of $45^\circ$, and} the lattice structure reaches the mechanical four-Weyl-line phase in Fig. \ref{fig4}\textbf{b}. Floppy modes arise on both the top and bottom surfaces, which is reflected by the comparable stiffness in Fig. \ref{fig4}\textbf{d}. Here, hysteresis in the force-displacement measurements stems from the combined effects of the hinge clearance and friction. We note that the fully-polarized topological and Weyl phases can be reversibly transformed by the uniform soft shearing strain.

The highly polarized and flexible boundary elasticity in 3D has significant implications, revolutionizing various applications. One such application involves switchable landing gear for drones (Fig. \ref{fig4}\textbf{e}). During landing, the material transitions into a soft-bottomed Weyl phase, effectively absorbing shocks. In flight, the material transforms into a topological phase with a rigid bottom surface, ensuring stability. Remarkably, these extraordinary functionalities persist even if the outer layers of the topological metamaterial are peeled off, showcasing its resilience and durability.

{Our pyrochlore structure applies to continuum materials, where finite bending stiffness raises the frequencies of topological floppy modes. These ``soft modes" propagate asymmetrically between the lattice’s top and bottom surfaces, leading to intriguing consequences. For example, our pyrochlore structure can be incorporated into a cylindrical domain, creating porous wheels that efficiently absorb energy on rugged terrains (see Figs. \ref{fig4}\textbf{f} and \textbf{g}).}

\emph{Discussions---}We have demonstrated, {both theoretically and experimentally,} the topologically fully-polarized mechanical phase in 3D. The mechanical metamaterial remains on the {isostatic} point, ensuring {the rigorousness of the topological mechanical index}. Topological floppy modes, confined exclusively to a single boundary, result in fully-polarized topological phase without Weyl lines. {This mechanical achievement is analogous to the discovery of 3D topological insulators in electronic systems.} Using {soft uniform shearing modes}, the {isostatic} metamaterial can be transformed from topologically polarized to Weyl phases, reducing the stiffness contrast between opposite boundaries to a trivially comparable level.

The topological polarization in 3D paves the way towards physics not possible for 2D lattices, such as higher-order topological floppy modes in 3D, topological mechanical cloaking, and static mechanical non-reciprocity in all spatial dimensions.

\emph{Acknowledgement---}D. Z., F. L., and Y. Y. acknowledge the support from the National Science Foundation of China (Grant Nos. 12374157, 12102039, 12272040, 11734003). 
D. Z. and F. L. acknowledge insightful discussions with Chiara Daraio and Ying Wu.

%
%
%
%
%
%
%
%
%
%
%
%
\begin{appendices}

\section{Experimental and numerical details of force-displacement measurements}

\begin{figure}[htbp]
	\includegraphics[scale=0.4]{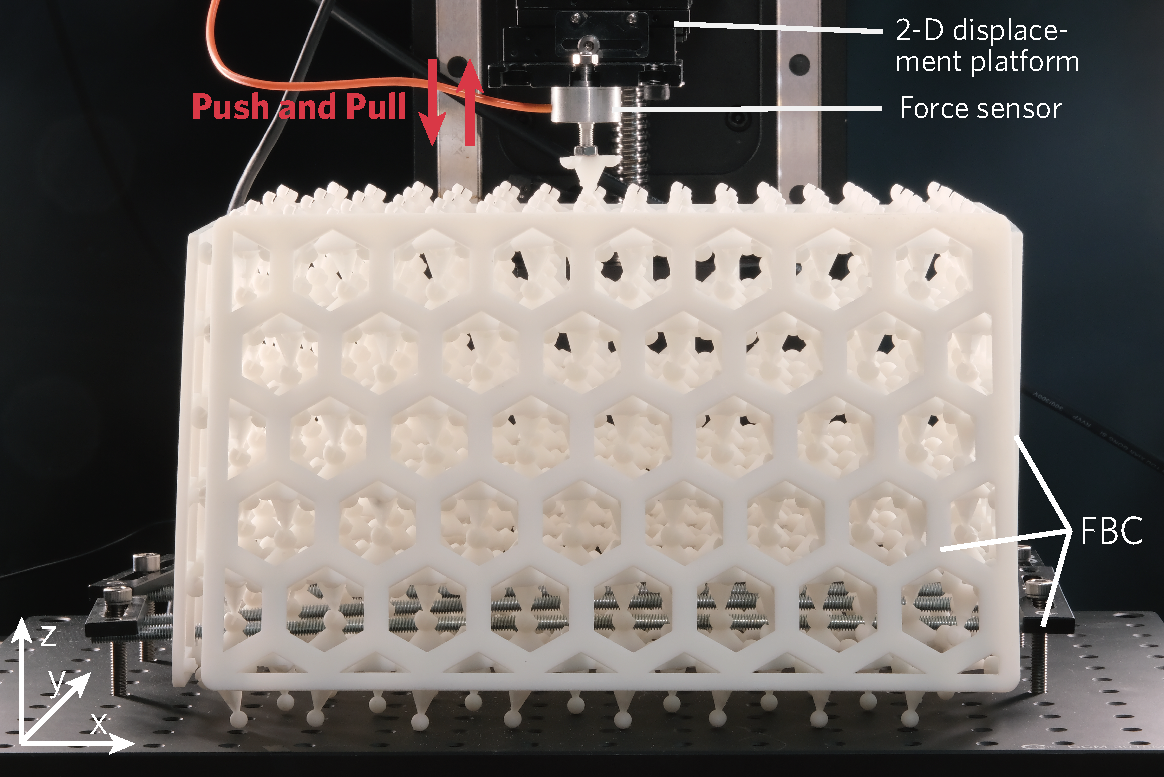}
	\caption{Experimental setup for the force-displacement measurement on top and bottom the open boundaries of the pyrochlore lattice.}\label{FigSIexp}
\end{figure}

We construct a generalized pyrochlore lattice that is consisted of $8\times 8\times 6$ unit cells. The top and bottom boundaries are left open, whereas four fixed boundary conditions are applied on all four lateral faces (front, back, left, right). These four fixed side boundaries (i.e., 4 FBCs) are consistent with those used in the experiment and the data presented in Figs. 4\textbf{c} and \textbf{d} of the main text.

To measure the stiffness of the top and bottom open boundaries, force sensors are placed on a two-dimensional displacement platform that allows free movement in the $xy$ plane. Our focus is primarily on studying the two open boundaries indicated by the $\bm{a}_3$ direction (i.e., open surfaces perpendicular to the reciprocal vector $\bm{b}_3$). Therefore, except for the one being measured, we ensure that all five boundaries (the bottom and the four side surfaces) are under fixed boundary conditions during measurements. The fixed-boundary constraints on the front and back boundaries are removed here for visual convenience of the internal structure of the lattice. By slowly and continuously pushing and pulling the measuring tip of the force sensor that is connected to the open boundary in the vertical direction, we obtain the force-displacement curves of the top and bottom open boundaries of the generalized pyrochlore lattice, as illustrated in Fig. 4\textbf{c} and \textbf{d} of the main text.

\begin{figure}[htbp]
	\includegraphics[scale=0.4]{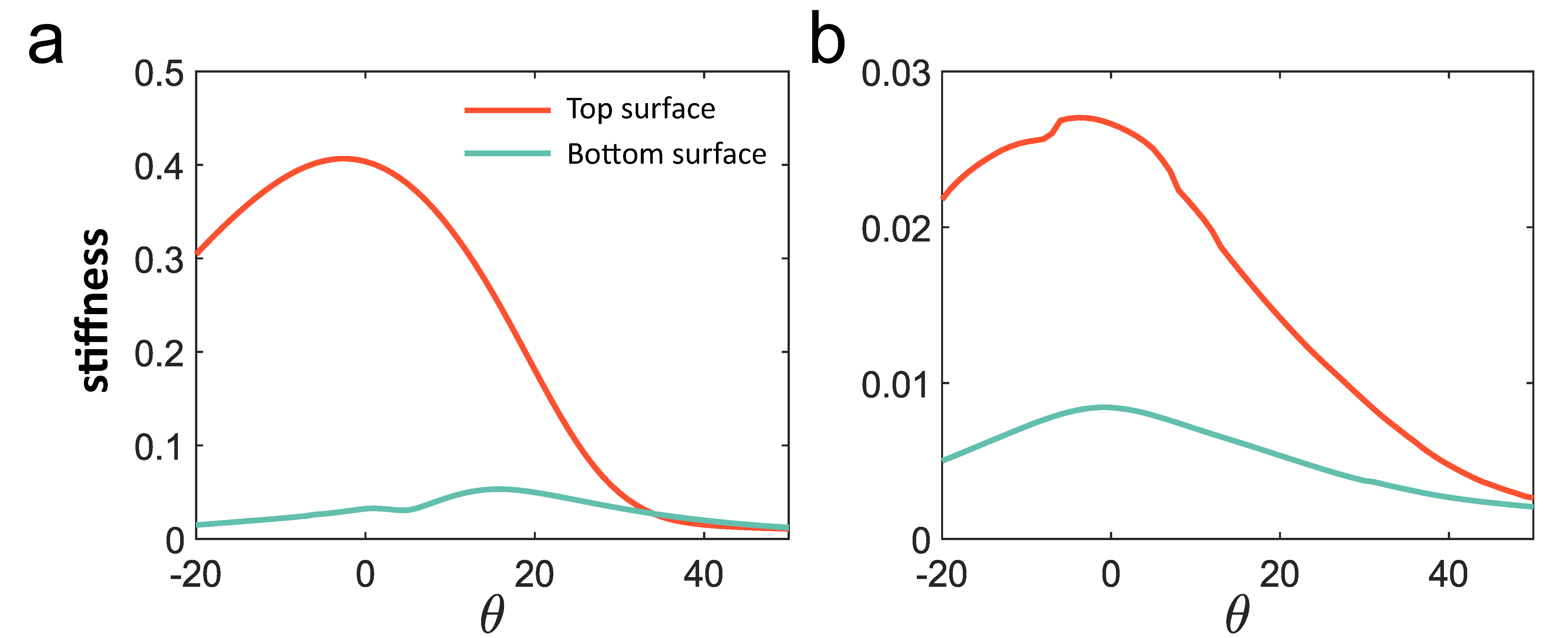}
	\caption{Numerical force-displacement measurement of the top and bottom open surfaces of the pyrochlore lattice. \textbf{a}: Fixed four side boundaries. \textbf{b}: Two fixed boundaries (left and right) and two open boundaries (front and back).}\label{figR10}
\end{figure}

Next, we ask on how much effect do the boundary conditions have on the stiffness contrast between the top and bottom open surfaces. To this end, we numerically construct the topologically polarized pyrochlore lattice that is composed of $5\times5\times8$ unit cells. We build two kinds of boundary conditions: One lattice uses fixed conditions for all four lateral sides (4 FBCs), which is consistent with the force-displacement measurement performed in the experiment, see SI. Fig. \ref{figR10}\textbf{a}; The other lattice uses open conditions for the front and back, and fixed conditions for the left and right (2 OBCs $+$ 2 FBCs), see SI. Fig. \ref{figR10}\textbf{b}. In the topologically polarized phase, the stiffness of the top and bottom boundaries under ``4 FBCs" in SI. Fig. \ref{figR10}\textbf{a} is significantly higher than that under ``2 OBCs$+$2 FBCs" in SI. Fig. \ref{figR10}\textbf{b}. This reduction in SI. Fig. \ref{figR10}\textbf{b} is mainly attributed to the additional floppy modes generated on the front and back open boundaries. Nevertheless, topological robustness in rigidity is retained on the upper surface, as evidenced by the significantly higher stiffness of the top boundary compared to the bottom boundary (see SI. Fig. \ref{figR10}\textbf{b}).

While the front and back open boundaries can reduce the stiffness of the top and bottom boundaries, which are physical properties in real space, these boundary conditions do not alter the topological polarization defined in reciprocal space. This is doubly confirmed by Kane-Lubensky theory~\cite{kane2014topological}, where the polarization reflects the intrinsic topological structure of the mechanical bulk bands defined in momentum space, and is therefore unaffected by real-space boundary conditions.

\section{Modeling the mechanics of the spring-mass model in the generalized pyrochlore lattice}

\begin{figure}[htbp]
	\includegraphics[scale=0.35]{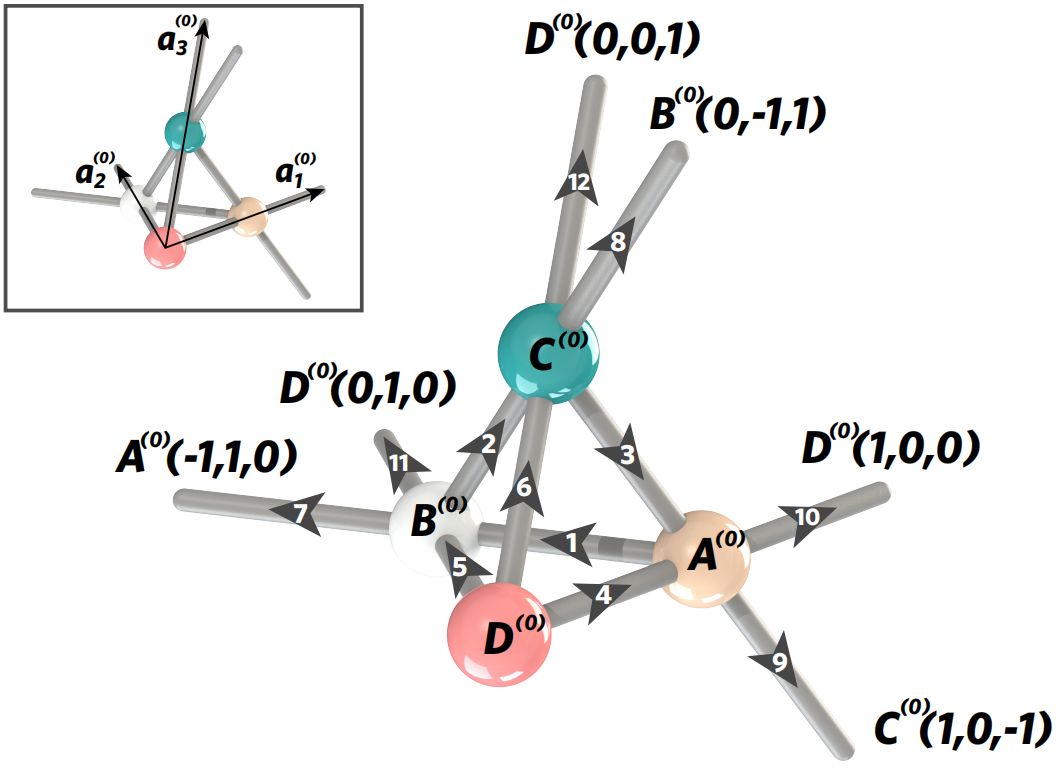}
	\caption{Representation of the unit cell geometry in the regular pyrochlore lattice. The primitive vectors of the lattice, denoted as $\bm{a}_1^{(0)}$, $\bm{a}_2^{(0)}$, and $\bm{a}_3^{(0)}$, define its structure. The unit cell contains four sites, marked as $A^{(0)}$, $B^{(0)}$, $C^{(0)}$, and $D^{(0)}$, which represent mass points. The Hookean bonds within the lattice are labeled from 1 to 12, and their orientations are indicated by the black arrows. The unit cell is conveniently represented by a vectorial index $\bm{n}=(n_1, n_2, n_3)$, where the sites are denoted as $A^{(0)}(\bm{n})$, $B^{(0)}(\bm{n})$, $C^{(0)}(\bm{n})$, and $D^{(0)}(\bm{n})$. To indicate sites connected to $A^{(0)}(\bm{n})$, $B^{(0)}(\bm{n})$, $C^{(0)}(\bm{n})$, and $D^{(0)}(\bm{n})$ but belong to other unit cells, they are labeled as $A^{(0)}(\bm{n}+(-1,1,0))$, $B^{(0)}(\bm{n}+(0,-1,1))$, $C^{(0)}(\bm{n}+(1,0,-1))$, $D^{(0)}(\bm{n}+(1,0,0))$, $D^{(0)}(\bm{n}+(0,1,0))$, and $D^{(0)}(\bm{n}+(0,0,1))$. For convenience, we simplify their notations by omitting the repeated index $\bm{n}$, resulting in $A^{(0)}(-1,1,0)$, $B^{(0)}(0,-1,1)$, $C^{(0)}(1,0,-1)$, $D^{(0)}(1,0,0)$, $D^{(0)}(0,1,0)$, and $D^{(0)}(0,0,1)$.}\label{SIfig1}
\end{figure}

This section aims to construct the mechanical properties of an idealized spring-mass model that is composed of mass particles connected by Hookean springs. The mass points are positioned at the vertices of a generalized pyrochlore lattice, and the elastic linear springs are arranged along the edges of this lattice. We analyze the mechanical properties of this spring-mass model by establishing the compatibility matrix. Finally, we derive the winding numbers that characterize the topological phases of the static mechanical properties of the generalized pyrochlore lattice. 

\subsection{The geometry of the regular pyrochlore lattice}

First of all, we define the geometry of the regular pyrochlore lattice, whose unit cell is illustrated in SI. Fig. \ref{SIfig1}. The unit cell is composed of four sites, denoted as $A$, $B$, $C$, and $D$, and 12 edges, labeled from 1 to 12. 

The site positions are defined at $A^{(0)} = \ell(1,1,0)/2$, $B^{(0)}= \ell(0,1,1)/2$, $C^{(0)} = \ell(1,0,1)/2$, and $D^{(0)} = \ell(0,0,0)$, where $\ell$ is the length scale of the idealized model, and the superscript ``$(0)$" implies that we are considering the geometry of the regular pyrochlore lattice. The primitive vectors, as indicated by the black arrows in the inset of SI. Fig. \ref{SIfig1}, are given by $\bm{a}^{(0)}_1 = \ell(1,1,0)$, $\bm{a}^{(0)}_2 = \ell(0,1,1)$, and $\bm{a}^{(0)}_3 = \ell(1,0,1)$. Thus, the unit cell can be labeled by three integer numbers $n_1$, $n_2$, and $n_3$, which indicates the position of the unit cell at $n_1\bm{a}_1^{(0)}+n_2\bm{a}_2^{(0)}+n_3\bm{a}_3^{(0)}$. In the rest of this Supplementary Information, we always use a vectorial index $\bm{n} = (n_1, n_2, n_3)$ to mark the unit cell. Following this convention, the sites are labeled by $A^{(0)}(\bm{n})$, $B^{(0)}(\bm{n})$, $C^{(0)}(\bm{n})$, and $D^{(0)}(\bm{n})$, whose positions are given by $X^{(0)}(\bm{n})=X^{(0)}+n_1\bm{a}_1^{(0)}+n_2\bm{a}_2^{(0)}+n_3\bm{a}_3^{(0)}$ for $X=A,B,C,D$.

Then, we elaborate on how nodes are connected in the network of regular pyrochlore lattice. To this end, we study the 12 edges within the unit cell, as shown in SI. Fig. \ref{SIfig1}, whose edge vectors are defined below, where again the superscript ``$(0)$" indicates that we are considering the regular pyrochlore lattice:
\begin{eqnarray}\label{A1}
	& {} & \bm{l}_1^{(0)} = B^{(0)}-A^{(0)},\qquad\qquad \bm{l}_2^{(0)} = C^{(0)}-B^{(0)},\nonumber \\
	& {} & \bm{l}_3^{(0)} = A^{(0)}-C^{(0)},\qquad\qquad
	\bm{l}_4^{(0)} = A^{(0)}-D^{(0)},\nonumber \\
	& {} & \bm{l}_5^{(0)} = B^{(0)}-D^{(0)},\qquad\qquad
	\bm{l}_6^{(0)} = C^{(0)}-D^{(0)},\nonumber \\
	& {} & \bm{l}_7^{(0)} = A^{(0)}-\bm{a}_1^{(0)}+\bm{a}_2^{(0)}-B^{(0)}, \nonumber \\
	& {} & \bm{l}_8^{(0)} = B^{(0)}-\bm{a}_2^{(0)}+\bm{a}_3^{(0)}-C^{(0)},\nonumber \\
	& {} & \bm{l}_9^{(0)} = C^{(0)}+\bm{a}_1^{(0)}-\bm{a}_3^{(0)}-A^{(0)},\nonumber \\
	& {} &  \bm{l}_{10}^{(0)} = D^{(0)}+\bm{a}_1^{(0)}-A^{(0)},\nonumber \\
	& {} & \bm{l}_{11}^{(0)} = D^{(0)}+\bm{a}_2^{(0)}-B^{(0)}, \nonumber \\
	& {} & \bm{l}_{12}^{(0)} = D^{(0)}+\bm{a}_3^{(0)}-C^{(0)}.
\end{eqnarray}

In SI. Eqs. (\ref{A1}), the edge vectors $\bm{l}_7^{(0)}$ through $\bm{l}_{12}^{(0)}$ include the primitive vectors $\bm{a}_1^{(0)}$, $\bm{a}_2^{(0)}$, and $\bm{a}_3^{(0)}$ in their expression because these edges connect sites that belong to different unit cells. For example, as shown in SI. Fig. \ref{SIfig1}, the edge vector $\bm{l}_{10}^{(0)}$ connects the sites $A^{(0)}(n_1,n_2,n_3)$ and $D^{(0)}(n_1+1,n_2,n_3)$, which belong to different unit cells labeled by $(n_1,n_2,n_3)$ and $(n_1+1,n_2,n_3)$. Thus, the corresponding edge vector is calculated as $\bm{l}_{10}^{(0)} = D^{(0)}(n_1+1,n_2,n_3)-A^{(0)}(n_1,n_2,n_3)$. The primitive vector $\bm{a}_1^{(0)}$ stems from the different unit cells that the sites $A^{(0)}(n_1,n_2,n_3)$ and $D^{(0)}(n_1+1,n_2,n_3)$ belong to. Likewise, the edge $\bm{l}_7^{(0)}$ connects the sites $B^{(0)}(n_1,n_2,n_3)$ and $A^{(0)}(n_1-1,n_2+1,n_3)$. Its edge vector is therefore given by $\bm{l}_7^{(0)} = A^{(0)}(n_1-1,n_2+1,n_3)-B^{(0)}(n_1,n_2,n_3)$.

In the regular pyrochlore lattice, all edges have the same length of $\ell/\sqrt{2}$. The corresponding unit vectors of these 12 edges are defined as 
\begin{eqnarray}
	\hat{t}_i^{(0)} = \bm{l}_i^{(0)}/|\bm{l}_i^{(0)}|,\qquad i=1,2,\ldots, 12.
\end{eqnarray}
For the regular pyrochlore lattice, we have $\hat{t}_1^{(0)}=(-1,0,1)/\sqrt{2}$, $\hat{t}_2^{(0)}=(1,-1,0)/\sqrt{2}$, $\hat{t}_3^{(0)}=(0,1,-1)/\sqrt{2}$, $\hat{t}_4^{(0)}=(1,1,0)/\sqrt{2}$, $\hat{t}_5^{(0)}=(0,1,1)/\sqrt{2}$, $\hat{t}_6^{(0)} = (1,0,1)/\sqrt{2}$. The other edges labeled from $7$ to $12$ are parallel to the edges labeled from $1$ to $6$, yielding $\hat{t}_{i+6}^{(0)}\parallel \hat{t}_{i}^{(0)}$ for $i=1,2,\ldots, 6$. This relationship does not hold for the generalized pyrochlore lattice, as we indicate below. 

\subsection{The geometry of the generalized pyrochlore lattice}

\begin{figure}[htbp]
	\includegraphics[scale=0.4]{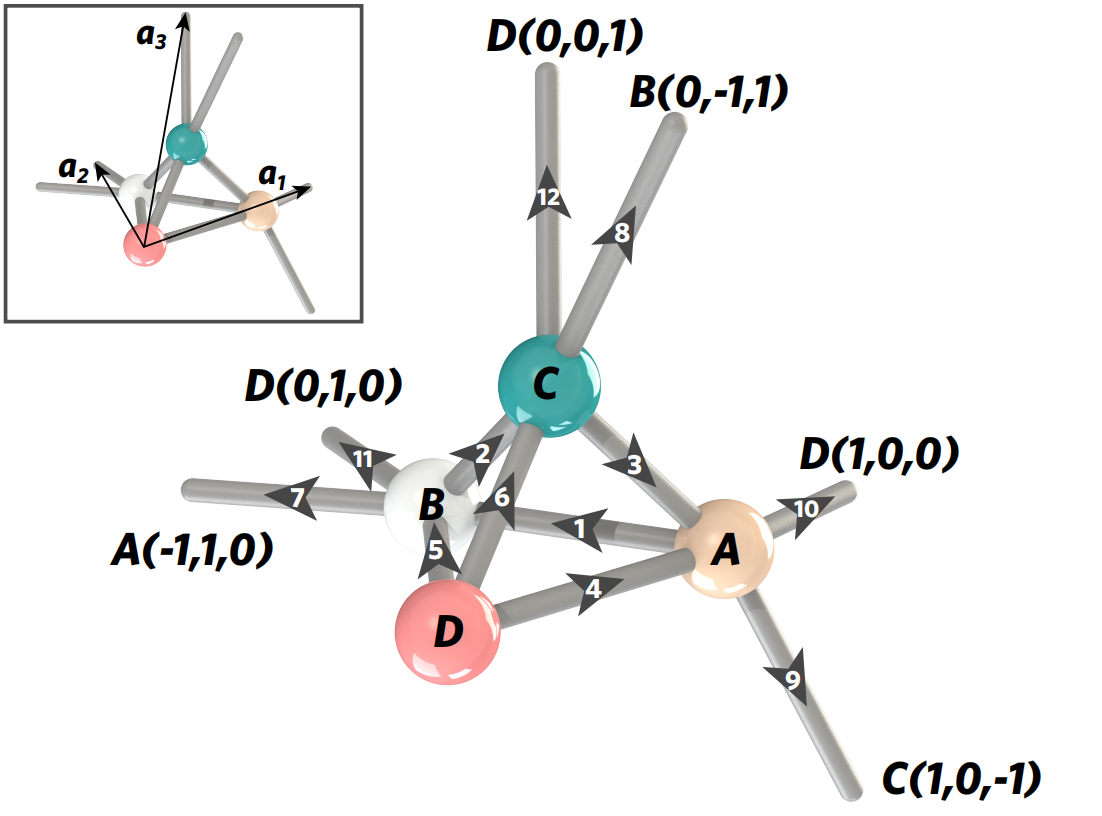}
	\caption{The geometry of the unit cell of the generalized pyrochlore lattice. $\bm{a}_1$, $\bm{a}_2$, and $\bm{a}_3$ denote the primitive vectors. $A$, $B$, $C$, and $D$ mark the four sites in the unit cell. The edges are labeled from 1 to 12, whose orientations follow the black arrows on top of them. }\label{SIfig2}
\end{figure}

The network connectivity of the generalized pyrochlore lattice, i.e., how nodes are connected to each other, remains the same as that of the regular pyrochlore lattice. However, the geometry of the generalized pyrochlore lattice, namely the primitive vectors, the site positions, and the edge vectors, are different from the regular pyrochlore lattice. 

Within the unit cell, the site positions are displaced from those of the regular lattice, as defined by $X=X^{(0)}+\Delta X$ for $X=A,B,C,D$, where $\Delta X$ denotes the change in node positions. The primitive vectors are defined by $\bm{a}_i=\bm{a}_i^{(0)}+\Delta \bm{a}_i$ for $i=1,2,3$, where $\Delta \bm{a}_i$ represents the difference of the primitive vectors between the generalized and regular pyrochlore lattices. The site positions in the generalized pyrochlore lattice are given by $X(\bm{n}) = X+n_1\bm{a}_1+n_2\bm{a}_2+n_3\bm{a}_3$, with $\bm{n}=(n_1, n_2, n_3)$ the vectorial index that labels the unit cell of the generalized pyrochlore lattice. Since the focus of this paper is to discuss the boundary mechanics of the pyrochlore structure under distinct geometries, we refer to such a class of pyrochlore lattice geometry that differs from the regular pyrochlore lattice, as the ``generalized pyrochlore lattice".


In SI. Sec. \uppercase\expandafter{\romannumeral2}, we will discuss the topologically polarized and Weyl phases of the static mechanical properties by specifying the geometric parameters of the generalized pyrochlore lattice.

The edge vectors in the generalized pyrochlore lattice follow the same definitions as the regular pyrochlore lattice. To be specific, the edge vectors are defined as follows,
\begin{eqnarray}\label{1}
	& {} & \bm{l}_1 = B-A,\qquad\qquad 
	\bm{l}_2 = C-B,\nonumber \\
	& {} & \bm{l}_3 = A-C,\qquad\qquad
	\bm{l}_4 = A-D,\nonumber \\
	& {} & \bm{l}_5 = B-D,\qquad\qquad
	\bm{l}_6 = C-D,\nonumber \\
	& {} & \bm{l}_7 = A-\bm{a}_1+\bm{a}_2-B, \nonumber \\
	& {} & \bm{l}_8 = B-\bm{a}_2+\bm{a}_3-C,\nonumber \\
	& {} & \bm{l}_9 = C+\bm{a}_1-\bm{a}_3-A,\nonumber \\
	& {} & \bm{l}_{10} = D+\bm{a}_1-A,\nonumber \\
	& {} & \bm{l}_{11} = D+\bm{a}_2-B, \nonumber \\
	& {} & \bm{l}_{12} = D+\bm{a}_3-C.
\end{eqnarray}
The edge orientations are defined as
\begin{eqnarray}\label{unitVec}
	\hat{t}_i = \bm{l}_i/|\bm{l}_i|,\qquad i=1,2,\ldots, 12. 
\end{eqnarray}
Finally, the site positions of the top tetrahedron within the unit cell are given by $A'=A-\bm{a}_1+\bm{a}_3$, $B'=B-\bm{a}_2+\bm{a}_3$, $C'=C$, and $D'=D+\bm{a}_3$. Consequently, the geometry of the generalized pyrochlore lattice, such as the site positions and edge vectors, are different from those of the regular pyrochlore lattice. Furthermore, as we indicated before, the edges for $i=7,8,\ldots,12$ are not parallel to the edges labeled by $i=1,2,\ldots,6$: $\hat{t}_{i+6}{\not \,\parallel}\,\hat{t}_i$ for $i=1,2,\ldots,6$. Finally, we define the reciprocal vectors of the generalized pyrochlore lattice as $\bm{b}_i = 2\pi\epsilon_{ijk} (\bm{a}_j\times \bm{a}_k) /[\bm{a}_1 \cdot(\bm{a}_2\times \bm{a}_3)]$ for $i,j,k=1,2,3$.

It is worth noting that, as long as the site positions $A$, $B$, $C$, and $D$ are given, the edge vectors from $\bm{l}_1$ to $\bm{l}_6$ of the bottom tetrahedron can be determined, as shown by SI. Eqs. (\ref{1}). Likewise, given the site positions of the bottom tetrahedron and the primitive vectors $\bm{a}_{i=1,2,3}$, all edge vectors of the top tetrahedron, from $\bm{l}_7$ to $\bm{l}_{12}$, are determined.

\subsection{Compatibility matrix of the static mechanics of the spring-mass pyrochlore lattice}

We consider a spring-mass model that is embedded in the geometry of the generalized pyrochlore lattice. As shown in SI. Fig. \ref{SIfig2}, each site is occupied by a particle with identical mass $m$. The edges are represented by central-force Hookean springs whose elastic constant are identically $k$, and their rest lengths match the corresponding edges of the generalized pyrochlore lattice. This is a three-dimensional mechanical model ($d=3$), whose unit cell possesses four mass points ($n_{\rm site}=4$). Therefore, a total of 12 degrees of freedom ($n_{\rm site} d = 12$) are contained in each unit cell. On the other hand, the unit cell has 12 Hookean springs to offer 12 constraints ($n_{\rm bond}=12$). As a result, the unit cell of the pyrochlore lattice has equal numbers of degrees of freedom and constraints ($n_{\rm site}  d =n_{\rm bond} =12 $), creating a lattice that is on the verge of mechanical instability. In other words, the inner body of the generalized pyrochlore lattice is at the isostatic point.

Having defined the ground state of the spring-mass pyrochlore model in which all bonds are un-stretched, we now consider particle displacements away from their equilibrium positions. Particle displacements induce elongations in the Hookean springs and cause elastic potential energy. We denote $\bm{u}_X(\bm{n})$ as the three-dimensional displacement of the site that is away from its equilibrium position, where $X=A,B,C,D$ marks the four sites within the unit cell, and $\bm{n}=(n_1,n_2,n_3)$ labels the associated unit cell. We consider particle displacements that are small comparing to the edge lengths, which allows for linear approximations in bond elongations:
\begin{eqnarray}\label{2}
	& {} & \delta l_1(\bm{n}) = [\bm{u}_B(\bm{n})-\bm{u}_A(\bm{n})]\cdot \hat{t}_1, \nonumber \\
	& {} & 
	\delta l_2(\bm{n}) = [\bm{u}_C(\bm{n})-\bm{u}_B(\bm{n})]\cdot\hat{t}_2,\nonumber \\
	& {} & \delta l_3(\bm{n}) = [\bm{u}_A(\bm{n})-\bm{u}_C(\bm{n})]\cdot\hat{t}_3,\nonumber \\
	& {} & 
	\delta l_4(\bm{n}) = [\bm{u}_A(\bm{n})-\bm{u}_D(\bm{n})]\cdot\hat{t}_4,\nonumber \\
	& {} & \delta l_5(\bm{n}) = [\bm{u}_B(\bm{n})-\bm{u}_D(\bm{n})]\cdot\hat{t}_5,\nonumber \\
	& {} & 
	\delta l_6(\bm{n}) = [\bm{u}_C(\bm{n})-\bm{u}_D(\bm{n})]\cdot\hat{t}_6,\nonumber \\
	& {} & \delta l_7(\bm{n}) = [\bm{u}_A(\bm{n}+(-1,1,0))-\bm{u}_B(\bm{n})]\cdot\hat{t}_7,\nonumber \\
	& {} & \delta l_8(\bm{n}) = [\bm{u}_B(\bm{n}+(0,-1,1))-\bm{u}_C(\bm{n})]\cdot\hat{t}_8,\nonumber \\
	& {} & \delta l_9(\bm{n}) = [\bm{u}_C(\bm{n}+(1,0,-1))-\bm{u}_A(\bm{n})]\cdot\hat{t}_9,\nonumber \\
	& {} & \delta l_{10}(\bm{n}) = [\bm{u}_D(\bm{n}+(1,0,0))-\bm{u}_A(\bm{n})]\cdot\hat{t}_{10},\nonumber \\
	& {} & \delta l_{11}(\bm{n}) = [\bm{u}_D(\bm{n}+(0,1,0))-\bm{u}_B(\bm{n})]\cdot\hat{t}_{11},\nonumber \\
	& {} & \delta l_{12}(\bm{n}) = [\bm{u}_D(\bm{n}+(0,0,1))-\bm{u}_C(\bm{n})]\cdot\hat{t}_{12},
\end{eqnarray}
where $\delta l_{i}(\bm{n})$ for $i=1,2,\ldots, 12$ stands for the change in the bond length of the $i$-th edge in the unit cell marked by $\bm{n}=(n_1, n_2, n_3)$.

Under mixed boundary conditions, such as periodic boundaries in the lattice directions $\bm{a}_1$ and $\bm{a}_2$, along with open boundaries in the lattice vector $\bm{a}_3$, the elastic bonds on the open boundaries need to be severed to create these open boundaries. As a result, the constraints and degrees of freedom on these open boundaries become unbalanced, leading to an excess of degrees of freedom in these conditions. Specifically, in a generalized pyrochlore lattice with mixed boundary conditions, denoted by the numbers of sites and bonds as $N_{\rm site}$ and $N_{\rm bond}$ respectively, the condition becomes $N_{\rm site}d > N_{\rm bond}$.

Furthermore, we denote the displacements of all particles as a $N_{\rm site}d$-dimensional vector field $\bm{u} = (\ldots, \bm{u}_A(\bm{n}),\bm{u}_B(\bm{n}),\bm{u}_C(\bm{n}),\bm{u}_D(\bm{n}),\ldots)^\top$ ($\top$ denotes matrix transpose), and denote the elongations of all bonds as a $N_{\rm bond}$-dimensional vector field $\delta \bm{l} = (\ldots, \delta l_1(\bm{n}), \delta l_2(\bm{n}), \ldots, \delta l_{12}(\bm{n}), \ldots)^\top$. Given the particle displacements $\bm{u}$, one can always find the bond elongations $\delta \bm{l}$ via SI. Eqs. (\ref{2}). Thus, SI. Eqs. (\ref{2}) allow us to establish a $N_{\rm bond}\times N_{\rm site}d$ matrix, namely the compatibility matrix $\mathbf{C}$~\cite{kane2014topological}, that maps particle displacements to bond elongations via $\mathbf{C}\bm{u} = \delta \bm{l}$. Compatibility matrix is useful in exploring both the \emph{dynamical and static properties} of the idealized spring-mass lattices, since the mechanical energy is governed by the compatibility matrix via $E = T+V = (m\dot{\bm{u}}^\top\dot{\bm{u}}+k\bm{u}^\top \mathbf{C}^\top \mathbf{C} \bm{u})/2$, where $T$ and $V$ are the kinetic and potential energy, respectively. Moreover, compatibility matrix can describe the \emph{static} mechanical properties of the 3D-printed pyrochlore metamaterial, as we have demonstrated in the main text. In what follows, we focus on the properties of this compatibility matrix.

In general, particle movements in the idealized spring-mass model induce both elastic deformation and kinetic energy. However, mechanical zero modes~\cite{lubensky2015RRP} refer to zero-frequency site displacements that do not deform elastic bonds. Therefore, all $\delta l_i(\bm{n})$ must be zero (i.e., $\delta \bm{l}=0$), resulting in zero potential energy $V=0$. Additionally, since mechanical zero modes occur very slowly, their zero-frequency static nature means that the contribution of kinetic energy of mass particles is also negligible, i.e., $T=0$. As a result, mechanical zero modes yield $\mathbf{C}\bm{u}=\delta\bm{l}=0$. According to linear algebra, the number of linearly independent mechanical zero modes corresponds to the dimensionality of the null space of the compatibility matrix, i.e., $N_{\rm zm} = \dim {\rm null}(\textbf{C})$. States of self-stress, however, describe non-zero tensions in elastic bonds that allow for vanishing net forces on every mass point. Thus, states of self-stress constitute the null space of the equilibrium matrix $\textbf{Q}$, where the equilibrium matrix $\textbf{Q}$ is equal to the transpose of the compatibility matrix $\textbf{C}^\top$. Consequently, the number of linearly independent states of self-stress equals to the dimensionality of the null space of the equilibrium matrix, $N_{\rm sss} = \dim{\rm null}(\textbf{Q})=\dim{\rm null}(\textbf{C}^\top)$. According to the rank-nullity theorem, which is also known as the Calladine-isostatic~\cite{lubensky2015RRP} theorem in mechanical lattices, the number difference between the mechanical zero modes and states of self-stress equals to the number difference between the degrees of freedom and constraints, 
\begin{eqnarray}
	N_{\rm zm}-N_{\rm sss} = N_{\rm site}d-N_{\rm bond}.
\end{eqnarray}

Floppy modes, on the other hand, are the (non-trivial) mechanical zero modes excluding the trivial translational and rotational zero modes, which amount to $d(d+1)/2=6$. Therefore, the number of linearly independent floppy modes is given by $N_{\rm fm}=N_{\rm zm}-d(d+1)/2$. Under open boundary conditions, the number of mechanical zero modes grows extensively as the length scale increases. Thus, for sufficiently large lattices, the number of floppy modes and mechanical zero modes,
\begin{eqnarray}
	N_{\rm fm}
	\approx N_{\rm zm},
\end{eqnarray}
are roughly the same, as the number contribution of trivial zero modes becomes negligible. Throughout the rest of this Supplementary Information, we treat the terminologies of floppy modes and mechanical zero modes interchangeably, without distinguishing between them. Below, we define winding numbers of isostatic lattices that characterize topological phases of these mechanical floppy modes in the generalized pyrochlore lattice.

\subsection{Topological winding numbers of the static mechanics in the spring-mass pyrochlore lattice}

Spatially repetitive frames allow for the representation of phonon modes as plane waves. These modes can be mathematically formulated using the Fourier transformation $\bm{u}_X(\bm{n}) = \sum_{\bm{k}}\bm{u}_X(\bm{k})e^{\mathrm{i}\bm{k}\cdot \bm{r}(\bm{n})}$. In this equation, $\bm{u}_X(\bm{n})$ represents the mechanical wave for the $X=A,B,C,D$ site in the unit cell, labeled by the vector index $\bm{n}=(n_1,n_2,n_3)$. The corresponding position 
is given by $\bm{r}(\bm{n})=n_1\bm{a}_1+n_2\bm{a}_2+n_3\bm{a}_3$, with $\bm{a}_1$, $\bm{a}_2$, and $\bm{a}_3$ the primitive vectors. $\bm{k}$ represents the phonon wavevector, and the summation $\sum_{\bm{k}}$ runs over the three-dimensional Brillouin zone. Furthermore, $\bm{u}_X(\bm{k})$ denotes the phonon mode at the wavevector $\bm{k}$, with $X=A,B,C,D$.

We denote $\bm{u}(\bm{k}) = (\bm{u}_A(\bm{k}), \bm{u}_B(\bm{k}), \bm{u}_C(\bm{k}), \bm{u}_D(\bm{k}))^\top$ as the $12\times 1$ displacement vector for the momentum $\bm{k}$. Likewise, the bond elongations can be decomposed as the plane-wave format via $\delta l_i(\bm{n}) = \sum_{\bm{k}}\delta l_i(\bm{k})e^{\mathrm{i}\bm{k}\cdot \bm{r}(\bm{n})}$ for $i=1,2,\ldots, 12$, where $\delta l_i(\bm{k})$ is the Fourier-transformed bond elongations at the momentum $\bm{k}$. Thus, we denote $\delta \bm{l}(\bm{k})=(\delta l_1(\bm{k}), \delta l_2(\bm{k}), \ldots, \delta l_{12}(\bm{k}))^\top$ as the $12\times 1$ column vector for the bond elongations of the wavevector $\bm{k}$. These Fourier-transformed quantities are related by the Fourier-transformed $12\times 12$ compatibility matrix $\mathbf{C}(\bm{k})$ via $\mathbf{C}(\bm{k})\bm{u}(\bm{k}) = \delta \bm{l}(\bm{k})$. For the generalized pyrochlore lattice, the Fourier-transformed compatibility matrix is elaborated as follows, 
\begin{eqnarray}\label{CM}
	& {} & \mathbf{C}(\bm{k})=\nonumber \\
	& {} & \left(
	\begin{array}{cccc}
		-\hat{t}_1 & \hat{t}_1 & 0 & 0\\
		0 & -\hat{t}_2 & \hat{t}_2 & 0\\
		\hat{t}_3 & 0 & -\hat{t}_3 & 0\\
		\hat{t}_4 & 0 & 0 & -\hat{t}_4\\
		0 & \hat{t}_5 & 0 & -\hat{t}_5\\
		0 & 0 & \hat{t}_6 & -\hat{t}_6\\
		e^{\mathrm{i}\bm{k}\cdot(\bm{a}_2-\bm{a}_1)}\hat{t}_7 & -\hat{t}_7 & 0 & 0\\
		0 & e^{\mathrm{i}\bm{k}\cdot (\bm{a}_3-\bm{a}_2)}\hat{t}_8 & -\hat{t}_8 & 0\\
		-\hat{t}_9 & 0 & e^{\mathrm{i}\bm{k}\cdot(\bm{a}_1-\bm{a}_3)}\hat{t}_9 & 0\\
		-\hat{t}_{10} & 0 & 0 & e^{\mathrm{i}\bm{k}\cdot \bm{a}_1}\hat{t}_{10}\\
		0 & -\hat{t}_{11} & 0 & e^{\mathrm{i}\bm{k}\cdot \bm{a}_2}\hat{t}_{11}\\
		0 & 0 & -\hat{t}_{12} & e^{\mathrm{i}\bm{k}\cdot \bm{a}_3}\hat{t}_{12}\\
	\end{array}
	\right).\nonumber \\
\end{eqnarray}

The compatibility matrix of the generalized pyrochlore lattice has a $14$-dimensional gigantic parameter space. As shown by Fig. 1\textbf{b} of the main text, the unit cell of the generalized pyrochlore lattice is composed of two tetrahedra that connect on site $C$. A total of $7d=21$ parameters in the compatibility matrix arise from the site positions $A$, $B$, $C$, and $D$, as well as the primitive vectors $\bm{a}_1$, $\bm{a}_2$, and $\bm{a}_3$. However, only 14 out of 21 parameters are freely adjustable. This can be understood by noticing that site $D$ can be translated to $D=(0,0,0)$ without affecting the mechanical properties of the pyrochlore lattice. Next, an overall rotation and an overall scaling of the entire structure will not modify the mechanical properties as the compatibility matrix only depends on the bond orientations. Thus, we fix the edge vector $\bm{l}_1=B-A$ as $(1,0,0)$, and ask the orientation of the $\Delta ABC$-triangle to lie in the horizontal plane. These constraints reduce the independent parameters to 14 in the compatibility matrix.

The topological properties of the floppy modes can be captured by the integer-valued winding numbers defined from the compatibility matrix in the reciprocal space. Given the wavevector $\bm{k}$ in the three-dimensional Brillouin zone, topological winding numbers are defined as follows, 
\begin{equation}\label{WN1}
	\mathcal{N}_i(\bm{k}) = -\frac{1}{2\pi\mathrm{i}}\oint_{\bm{k}\to \bm{k}+\bm{b}_i} d\bm{k} \cdot \nabla_{\bm{k}}\ln \det \mathbf{C}(\bm{k}),
\end{equation}
where $i=1,2,3$ labels the three reciprocal vector directions, $\bm{b}_i$ is the $i$-th reciprocal vector, and the integration trajectory, $\bm{k}\to \bm{k}+\bm{b}_i$, follows a straight and closed loop that is parallel to $\bm{b}_i$. For each $\bm{k}$, there are a total of three topological winding numbers in the generalized pyrochlore lattice.

In our numerical computation, the evaluation of winding numbers is performed by the following format, 
\begin{eqnarray}\label{wind1}
	\mathcal{N}_i(\bm{k}) & = & \frac{1}{2\pi}\sum_{n=1}^N |f_n|^{-2} \big[ {\rm Im}\, f_n\,\, {\rm Re}\,(f_{n+1}-f_n)\nonumber \\
	& {} & - {\rm Re}\, f_n\,\, {\rm Im}\,(f_{n+1}-f_n) \big],
\end{eqnarray}
where $f_n=\det \mathbf{C}\left(\bm{k}+\frac{n}{N}\bm{b}_i\right)$ is the determinant of the compatibility matrix at wavevector $\bm{k}+\frac{n}{N}\bm{b}_i$, $\bm{k} = \frac{n_j}{N'}\bm{b}_j+\frac{n_k}{N'}\bm{b}_k$ is the starting wavevector of the integration, $i,j,k=1,2,3$, $N=2000$ steps is used in the integration, $N'=1000$ is used, and we run over $1\le n_1,n_2\le N'$ to scan the winding numbers in the Brillouin zone.

These three numbers together govern the topological mechanical phase of isostatic lattices, as we will address in SI. Sec. \uppercase\expandafter{\romannumeral2}.

\subsection{Guest-Hutchinson modes}

Isostatic lattices can host nonlinear and uniform soft strains of the whole structure, known as Guest-Hutchinson modes~\cite{Guest2003JMPS}, that reversibly evolve the lattice geometry without causing any elongations of the edges. The 3D spring-mass model integrated into the geometry of the generalized pyrochlore lattice facilitates the Guest-Hutchinson modes, which are characterized by the relative rotations on the spherical hinge that connects the top and bottom tetrahedra within the unit cell, as shown by Fig. 1\textbf{b} of the main text. Therefore, Guest-Hutchinson modes can be quantified by the three Euler angles in the generalized pyrochlore lattice. This can be visualized in Fig. 2\textbf{a-c} and Fig. 4\textbf{a} and \textbf{b} of the main text, where we choose a Guest-Hutchinson mode shearing that rotates about the dashed-line axis on the spherical hinge.

By manipulating the bond orientations in the generalized pyrochlore lattice using this shearing Guest-Hutchinson mode, the compatibility matrix, as described in SI. Eq. (\ref{CM}), change accordingly. The topological winding numbers of the static mechanical properties, as shown in SI. Eq. (\ref{WN1}), can change as well. This shearing Guest-Hutchinson mode allows the pyrochlore lattice to achieve multiple topological phases of the mechanical properties, which we address below.

To facilitate a rich topological phase transition within the metamaterial, encompassing topologically polarized, two-Weyl-line, and four-Weyl-line phases, we follow two key considerations of the spherical hinges: (1) Opening angle of spherical hinges: We carefully selected the opening angle of the spherical hinges to allow a large tunable range among topologically distinct phases. Simultaneously, we ensured that the concave hinge fits into the convex side, preventing separation. Consequently, the opening angle of the concave spherical hinge is set at 110 degrees. (2) Central orientation of spherical hinges: By appropriately orienting the central axes of the concave spherical hinges, we enable a rich transformation from the initial topologically polarized phase to the four-Weyl-line phase after a 45-degree Guest-Hutchinson mode rotation.

\section{Topological phases of the static mechanics in the generalized pyrochlore lattice}


In this section, we will introduce the notions of fully-polarized topological phases, partially-polarized topological phases, and Weyl phases by focusing on the winding numbers. Furthermore, we will discuss the non-trivial mechanical properties of the corresponding isostatic lattices.


\subsection{Fully-polarized topological mechanical phases}

\begin{figure}[htbp]
	\includegraphics[scale=0.4]{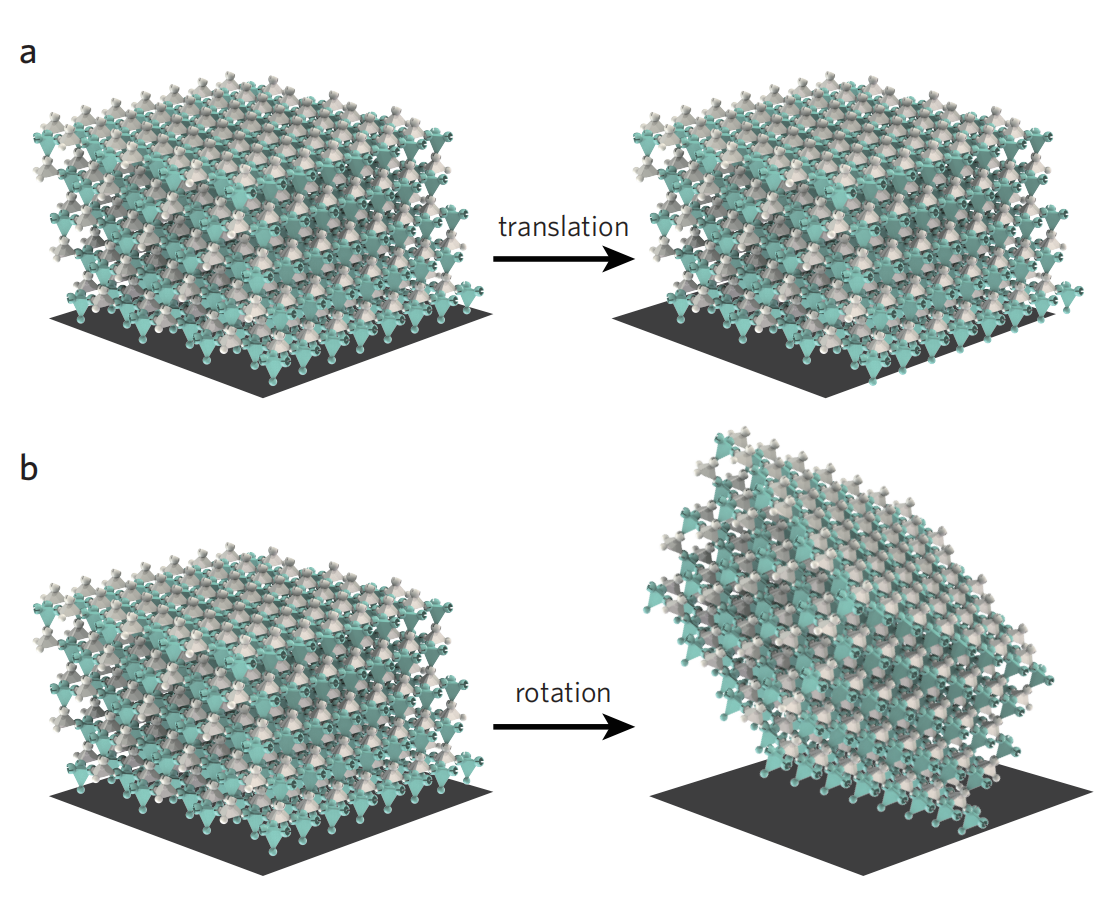}
	\caption{Translational and rotational mechanical zero modes in \textbf{a} and \textbf{b}, respectively.}\label{figR11}
\end{figure}

For certain geometric configurations of the generalized pyrochlore lattice, such as the unit cell configuration displayed in Fig. 1\textbf{b} of the main text, the phonon band structure can be gapped throughout the three-dimensional Brillouin zone, except for the trivial $\bm{k}=0$ point. This $\bm{k}=0$ point represents the wavevector of mechanical zero modes that translate or rotate the entire lattice as a whole, and these modes are commonly referred to as ``trivial mechanical zero modes"~\cite{lubensky2015RRP}. SI. Fig. \ref{figR11} provides a visual representation of how the lattice displaces under the translational or rotational mechanical zero modes.  

This gapped mechanical phase can be realized in the generalized pyrochlore lattice with the following set of geometric parameters
\begin{eqnarray}\label{deform1}
	& {} & \Delta A=0.053(1,0.3,0)\ell, \nonumber \\
	& {} & \Delta B=0.053(0.55,0,1)\ell, \nonumber \\ 
	& {} & \Delta C=0.053(-1,1,-1)\ell, \nonumber \\
	& {} & \Delta D=0.053(-1.8,-1,1.2)\ell, \nonumber \\
	& {} & \Delta\bm{a}_1=\Delta\bm{a}_2=\Delta\bm{a}_3=0.
\end{eqnarray}
Such a set of parameters has been used in the main text Fig. 1\textbf{b} and Fig. 2\textbf{a}. As a result, all winding numbers, i.e., $(\mathcal{N}_1(\bm{k}),\mathcal{N}_2(\bm{k}),\mathcal{N}_3(\bm{k}))=(1,-1,-2)$, stay invariant for arbitrary $\bm{k}$ in 3D reciprocal space, as we show the numbers in Figs. 2\textbf{a} and \textbf{d} using the numerical computation presented in SI. Eq. (\ref{wind1}). Furthermore, these \emph{invariant} integer numbers can be combined into a topologically invariant vector, called the topological polarization, 
\begin{eqnarray}
	\bm{R}_{\rm T}=\sum_{i=1}^3 \mathcal{N}_i \bm{a}_i.
\end{eqnarray}
For the geometric configuration presented in SI. Eq. (\ref{deform1}) (Fig. 2\textbf{a} of the main text), the topological polarization reads $\bm{R}_{\rm T}=\bm{a}_1-\bm{a}_2-2\bm{a_3}$. The physical significance of this topological polarization is that the topological floppy modes localize preferably on the surface that the topological polarization points to, whereas the parallel opposite surface has fewer topological mechanical floppy modes. In other words, $\bm{R}_{\rm T}$ indicates the topological polarization of the boundary floppy modes in the gapped phase of the considered isostatic lattice.

The principle of bulk-boundary correspondence states that the topological polarization obtained from the bulk phonon bands plays a crucial role in determining the number density of mechanical floppy modes on open surfaces of isostatic lattices. However, the local details of how nodes and bonds are connected on the open surfaces also play a role in determining the distribution of floppy modes. This local connectivity is quantitatively captured by the local polarization vector $\bm{R}_{\rm L}$. Together, topological and local polarizations govern the number distribution of boundary floppy modes via the relationship
\begin{equation}\label{density}
	\nu=\frac{1}{2\pi}\bm{G}\cdot(\bm{R}_{\rm T}+\bm{R}_{\rm L}),
\end{equation}
where the reciprocal lattice vector $\bm{G}$ is perpendicular to the considered open surface and points outward. In Fig. 1\textbf{d} and Fig. 2\textbf{a}, the reciprocal vectors for the top and bottom boundaries that are perpendicular to $\bm{b}_3$, are $\bm{G}_\uparrow = +\bm{b}_3$ and $\bm{G}_\downarrow = -\bm{b}_3$, respectively. The local polarizations are $\bm{R}_{\rm L\uparrow}=-\bm{a}_2+2\bm{a}_3$ and $\bm{R}_{\rm L\downarrow}=\bm{a}_1-\bm{a}_3$ for the top and bottom open surfaces, respectively. As a result, the number densities of surface floppy modes per supercell are $\nu_\uparrow=\bm{b}_3\cdot(\bm{R}_{\rm T}+\bm{R}_{\rm L\uparrow})/2\pi=0$ and $\nu_\downarrow=-\bm{b}_3\cdot(\bm{R}_{\rm T}+\bm{R}_{\rm L\downarrow})/2\pi=3$ for the top and bottom surfaces.

The floppy mode distribution, $\nu_\uparrow=0$, indicates that the top surface of the generalized pyrochlore lattice is completely devoid of mechanical floppy modes that account for the softness, rendering the open surface as rigid as the inner body of the lattice. In contrast, the floppy mode distribution, $\nu_\downarrow=3$, exhibits a much softer local response due to the exclusive emergence of topological floppy modes on the bottom boundary. This conclusion, as indicated from the perspective of momentum space, can be alternatively demonstrated from the perspective of real-space analysis. As shown in Fig. 3 of the main text, we perform a real-space analysis on calculating the numbers of floppy modes on a pair of parallel open surfaces normal to $\bm{b}_3$. The highly polarized distribution of floppy modes on the bottom boundary is in agreement with the result indicated by the topological polarization. 

Since the top surface is completely devoid of mechanical floppy modes, as indicated by $\nu_\uparrow = 0$, we dub the topological mechanical phase of this generalized pyrochlore lattice as the ``fully-polarized topological phase". This highly asymmetric boundary mechanics is topologically protected, in the sense that even if a few lattice layers are peeled off, the polarized mechanics remains unchanged. For instance, new top (bottom) surfaces may become rigid (soft) when the outer layers peel off due to severe conditions. The only way to modify the edge properties is to \emph{globally} alter the band topology of the phonon spectrum through a uniform twisting, referred to as Guest-Hutchinson modes, which we address in the following section.

\subsection{Partially-polarized topological mechanical phases}

Mechanically isostatic lattices with gapped phonon spectrum can manifest globally defined winding numbers and topological polarization vector. The number densities of boundary floppy modes follow SI. Eq. (\ref{density}). In the aforementioned section, we design the geometry of the pyrochlore lattice such that the top boundary is completely devoid of floppy modes, and all modes are localized on the bottom boundary. This highly polarized floppy mode localization leads to the strongly contrasting surface stiffness. However, in certain geometries of the generalized pyrochlore lattices, both of the parallel and opposite open boundaries can arise topological edge floppy modes despite the gapped phonon band structure and the well-defined topological polarization vector. Such geometric configuration has been illustrated in Ref.~\cite{Olaf2016PRL}, in which the bottom boundary hosts $\nu_\downarrow=2$ floppy modes per supercell, and the top surface hosts $\nu_\uparrow=1$ floppy mode. In this case, the stiffness contrast between the top and bottom open boundaries is in a trivially comparable level, which we refer to as ``partially-polarized topological mechanical phase".

\subsection{Weyl phase with mechanical Weyl lines}

In the previous section, the phonon band structure is gapped everywhere except for the trivial $\bm{k}=0$ point. This ``gapped phase" is obtained from the geometric configuration elaborated in SI. Eq. (\ref{deform1}). In this configuration, all winding numbers, $\mathcal{N}_{i=1,2,3}(\bm{k})$, remain invariant for arbitrary $\bm{k}$ in the momentum space.

However, in most geometries of generalized pyrochlore lattices, the phonon band structure can be gapless at some wavevectors $\bm{k}\neq 0$ in the three-dimensional Brillouin zone. For example, using the Guest-Hutchinson mode in SI. Sec. \uppercase\expandafter{\romannumeral1}(E) that uniformly shears the top tetrahedra relative to the mutual spherical hinges on the bottom tetrahedra throughout the entire lattice, the band structure of the isostatic pyrochlore lattice can be transformed in such a gapless state. To be specific, starting from the lattice configuration in Figs. 1\textbf{b} and 2\textbf{a} of the main text, we shear the top tetrahedron relative to the bottom one for every unit cell, with respect to the axis $\bm{a}_1\times\bm{b}_3$ with a counterclockwise rotation angle of $10^\circ$. In this $10^\circ$-sheared configuration, $\Delta A$, $\Delta B$, $\Delta C$, and $\Delta D$ remain the same as those in SI. Eqs. (\ref{deform1}), whereas the change of the primitive vectors, namely $\Delta\bm{a}_{i=1,2,3}$, are given by  
\begin{eqnarray}\label{deform2}
	& {} & \Delta\bm{a}_1=(0.075,-0.0966,0.0862)\ell, \nonumber \\
	& {} & \Delta\bm{a}_2=(0.0383,-0.0572,0.0477)\ell, \nonumber \\
	& {} & \Delta\bm{a}_3=(-0.0212,-0.1049,0.0418)\ell. 
\end{eqnarray}
This sheared configuration has a phonon band structure that is gapless at finite wavevectors. These band-touching points at finite wavevectors correspond to topologically protected bulk floppy modes that extend throughout the frame, and are called ``mechanical Weyl modes". The wavevectors of Weyl modes further constitute gapless lines in the three-dimensional Brillouin zone, which are referred to as ``mechanical Weyl lines". Isostatic lattices that host Weyl lines in their Brillouin zones are known to lie in the mechanical Weyl phases.

\begin{figure}[htbp]
	\includegraphics[scale=0.29]{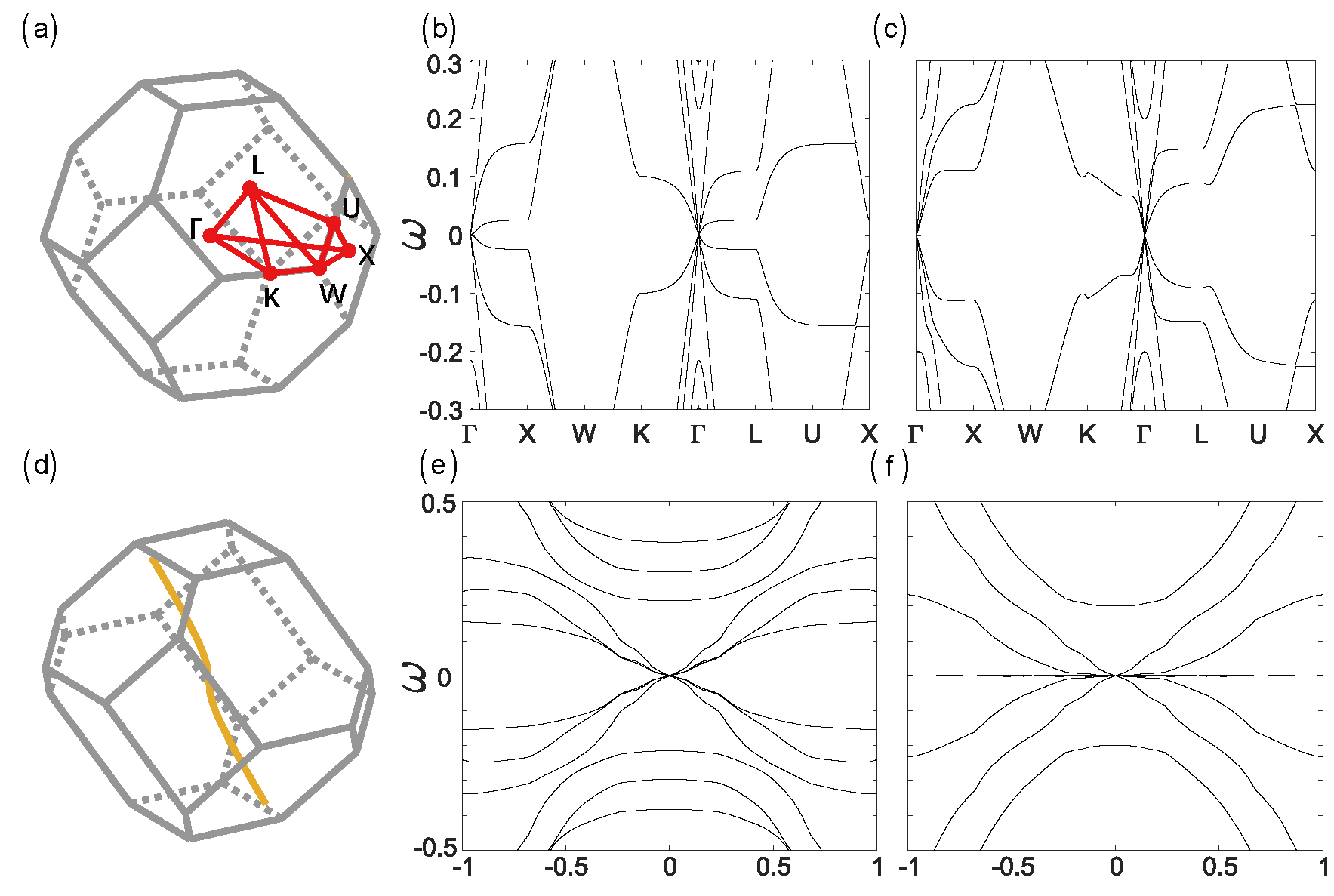}
	\caption{Mechanical band structures of generalized pyrochlore lattices. \textbf{a} shows high-symmetry lines in the Brillouin zone. \textbf{b} and \textbf{c} display the mechanical band structures for the topologically polarized and Weyl phases, respectively, along the high-symmetry lines. \textbf{d} defines a mechanical Weyl line in the Brillouin zone of the Weyl phase. \textbf{e} and \textbf{f} provide computations of the mechanical band structures for the topologically polarized and Weyl phases, respectively, where the wavevectors follow the trajectory of the mechanical Weyl line defined in \textbf{d}.}\label{figR12}
\end{figure}

To quantitatively discuss the aforementioned ``gapped" and ``gapless" mechanical band structures in the topologically polarized and Weyl phases, respectively, we compute the eigenvalues of the mechanical Hamiltonian for isostatic lattices. This Hamiltonian is obtained by taking the ``square root" of the Newtonian dynamical matrix~\cite{kane2014topological}. By doing so, we arrive at a Schr\"{o}edinger-like mechanical Hamiltonian for the pyrochlore lattice, 
\begin{eqnarray}\label{Hk}
	\textbf{H}(\bm{k}) = \left(\begin{array}{cc}
		0 & \textbf{C}(\bm{k})\\
		\textbf{C}^\dag(\bm{k}) & 0\\
	\end{array}\right)
\end{eqnarray}
where $\textbf{C}(\bm{k})$ is referred to as the compatibility matrix in reciprocal space. The mechanical band structures are numerically computed by finding the eigenvalues of $\textbf{H}(\bm{k})$.

SI. Figs. \ref{figR12}\textbf{a} to \textbf{f} present the mechanical band structures for the pyrochlore lattices in the topologically polarized and Weyl phases. As displayed in SI. Fig. \ref{figR12}\textbf{a}, we define the high-symmetry lines in the Brillouin zone. Following these lines, SI. Figs. \ref{figR12}\textbf{b} and \textbf{c} compute the mechanical band structures for the topologically polarized and Weyl phases, respectively. These bands are gapped for the wavevectors along the high-symmetry lines except for the $\Gamma$-point, where the $\bm{k}=0$ wavevector corresponds to the trivial mechanical zero modes that translationally or rotationally displace the lattice, as previously illustrated in SI. Fig. \ref{figR11}.

We further find that, interestingly, for certain wavevector trajectory, the behaviors of the corresponding mechanical bands along this wavevector trajectory are different for the topologically polarized and Weyl phases. This wavevector trajectory is defined in SI. Fig. \ref{figR12}\textbf{d}, where the wavevectors follow a Weyl line of the mechanical Weyl phase. By following this wavevector trajectory, we compute the mechanical band structure in the topologically polarized phase, which still reflects the ``gapped nature everywhere except for the trivial  point" (SI. Fig. \ref{figR12}\textbf{e}). However, in the Weyl phase, the mechanical band structure displays gaplessness across all wavevectors along this trajectory, due to the emergence of zero-frequency Weyl bulk states (SI. Fig. \ref{figR12}\textbf{f}).

\begin{figure}[htbp]
	\includegraphics[scale=0.4]{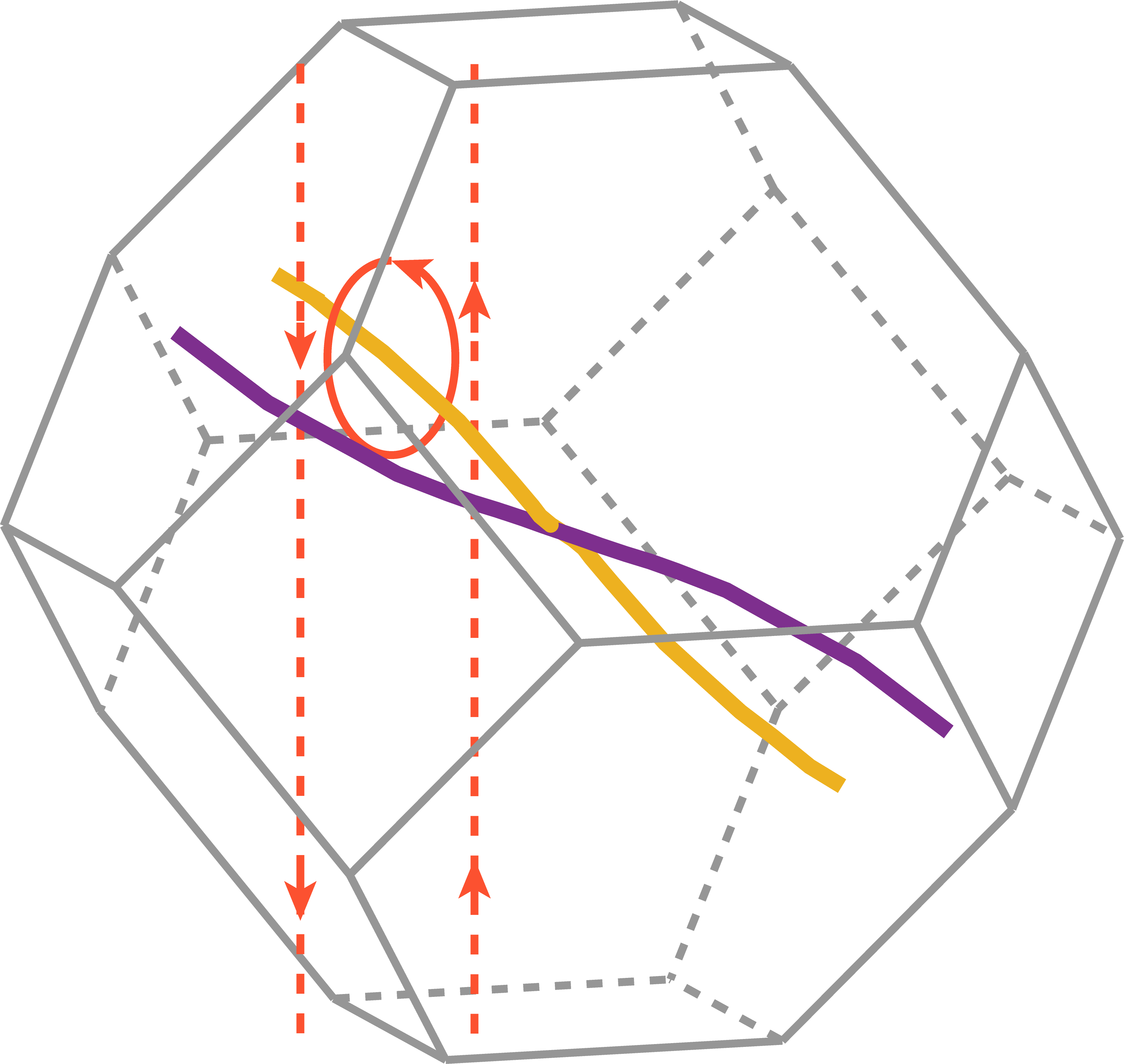}
	\caption{Two topologically protected Weyl lines, marked in orange and purple, are observed in the three-dimensional Brillouin zone. These Weyl lines correspond to the Guest-Hutchinson mode angle of $\theta=10^\circ$ in Fig. 2\textbf{c} in the main text. To determine the topological nature of the orange Weyl line, we integrate the Berry phase of the compatibility matrix around a contour, depicted as the solid red circle, which encloses the orange Weyl loop. The resulting winding number, an integer value, is used to verify the topological charge of the orange Weyl line. It is worth noting that the contour can be deformed into a pair of red dashed lines that travel in opposite directions. This deformation showcases the change in the winding number across the Brillouin zone as the wavevector moves around.}\label{SIfig3}
\end{figure}

Mechanically isostatic lattices, such as the generalized pyrochlore lattice, can have multiple pairs of gapless Weyl lines depending on its geometric configuration. For instance, in the ``10$^\circ$-sheared" generalized pyrochlore lattice (see corresponding unit cell configuration in the inset of Fig. 2\textbf{c} and the specific parameters in SI. Eqs. (\ref{deform2})), the lattice has one pair of Weyl lines in the reciprocal space, as shown in SI. Fig. \ref{SIfig3}. 



Weyl lines have a significant impact on the topological winding numbers as well as the phases of the static mechanical properties in isostatic lattices. First, winding number $\mathcal{N}_i(\bm{k})$ for $i=1,2,3$ are no longer globally defined in the Brillouin zone. They change from one integer to another as the integration trajectory $\bm{k}\to\bm{k}+\bm{b}_i$ moves from the exterior to the interior of the Weyl lines, as shown by the red dashed lines in SI. Fig. \ref{SIfig3}. Second, winding numbers become ill-defined when their integration trajectory intersects with Weyl lines, because the compatibility matrix becomes singular at these gapless wavevectors. Thus, we seek for an index that quantitatively characterizes the topological charge of gapless Weyl lines. This topological number is depicted by the following Berry phase,
\begin{equation}\label{WN2}
	\mathcal{N}_w = -\frac{1}{2\pi\mathrm{i}}\oint_{C} d\bm{k} \cdot \nabla_{\bm{k}}\ln \det \mathbf{C}(\bm{k}),
\end{equation}
whose integration trajectory $C$ is a closed loop and encloses the considered Weyl line, as illustrated by the red circle in SI. Fig. \ref{SIfig3}. This integer value, as shown in SI. Eq. (\ref{WN2}), describes the topological nature of the mechanical Weyl line in reciprocal space, which corresponds to the jump in the winding number in SI. Eq. (\ref{WN1}) when the integration trajectory moves from the outside to the inside of the Weyl line. 

Since the topological charge of Weyl lines cannot change continuously, they are topologically protected against variations in geometric parameters of the pyrochlore lattice. As the total topological charge of Weyl lines are neutral, Weyl lines always emerge in pairs. This can be alternatively understood by noticing the time-reversal-symmetry in Newtonian mechanics, where every gapless wavevector $\bm{k}$ indicates a partner gapless wavevector at $-\bm{k}$. Thus, mechanical Weyl lines are topologically robust and cannot be easily removed until pairs of Weyl lines meet and annihilate, which also reflects the topological robustness of Weyl lines. This property poses the major challenge in realizing mechanical metamaterials clear of Weyl lines in the precedented works~\cite{Olaf2016PRL, Huber2017AM, Vitelli2017PNAS}, which we have realized previously in SI. Sec. \uppercase\expandafter{\romannumeral2}(A).

Mechanically isostatic lattices can have more than one pair of gapless Weyl lines in their reciprocal spaces, depending on their geometric configurations. For example, 
We further shear the lattice configuration from the configuration in SI. Eq. (\ref{deform2}) with the Guest-Hutchinson mode angle of $35^\circ$. The new geometry of the generalized pyrochlore lattice has $\Delta A$, $\Delta B$, $\Delta C$, and $\Delta D$, remain the same as those in SI. Eq. (\ref{deform1}), whereas the change in the primitive vectors, $\Delta\bm{a}_1$, $\Delta\bm{a}_2$, and $\Delta\bm{a}_3$, are given by 
\begin{eqnarray}\label{deform3}
	& {} & \Delta\bm{a}_1 = (1.1670,0.4673,0.3499)\ell,\nonumber \\
	& {} & \Delta\bm{a}_2 = (0.0728,0.6935,1.1896)\ell,\nonumber \\
	& {} & \Delta\bm{a}_3 = (0.7755,-0.4388,1.1071)\ell.
\end{eqnarray}
This configuration is pictorially depicted in the inset of Fig. 2\textbf{e}, and can be equivalently understood by shearing the unit cell configuration in the inset of Fig. 2\textbf{a} with a Guest-Hutchinson mode angle of $45^\circ$ (i.e., $10^\circ+35^\circ$). This configuration is also in the mechanical Weyl phase, but a total of four Weyl lines emerge in the Brillouin zone. While both the lattice structures in Figs. 2\textbf{c} and 2\textbf{e} are in the Weyl phases, these two Weyl phases are topologically distinct as they have different numbers of mechanical Weyl lines. The generalized pyrochlore lattice experiences a topological phase transition, from the two-Weyl-line phase in Figs. 2\textbf{c} to the four-Weyl-line phase in Figs. 2\textbf{e}. 


\subsection{Bulk-boundary correspondence of topological floppy modes in isostatic lattices}

In this section, we demonstrate the bulk-boundary correspondence by relating winding numbers to the number of localized floppy modes on open boundaries of isostatic lattices. To this end, we first impose periodic boundary conditions to the generalized pyrochlore lattice in all three lattice vector directions. This allows the parameterization of the compatibility matrix in terms of the wavevector in the three-dimensional Brillouin zone, through $\textbf{C} = \textbf{C}(\bm{k})$, where the matrix elements are itemized in SI. Eq. (\ref{CM}). Likewise, floppy modes $\bm{u}(\bm{k})$ can be parameterized by the wavevectors $\bm{k}$ as well, and they yield the defining equality $\textbf{C}(\bm{k})\bm{u}(\bm{k})=0$. The non-vanishing solution of $\bm{u}(\bm{k})$ demands the compatibility matrix to satisfy
\begin{eqnarray}\label{det}
	\det\textbf{C}(\bm{k})=0.
\end{eqnarray}
Under periodic boundary conditions, the numbers of floppy modes and states of self-stress are both zero, implying that SI. Eq. (\ref{det}) does not have real-valued $\bm{k}$ solution.

\begin{figure}[htb]
	\includegraphics[scale=0.4]{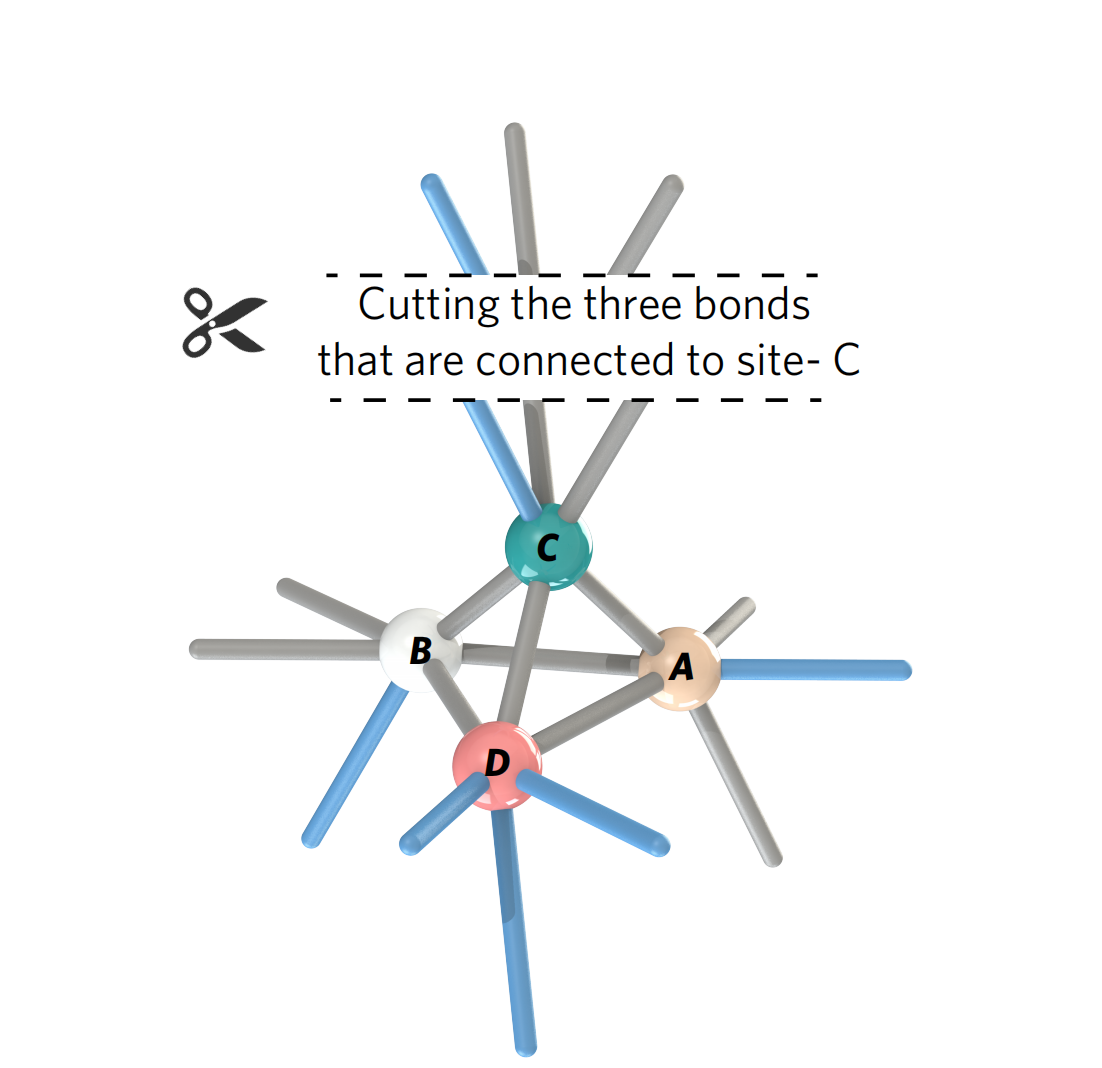}
	\caption{Cutting the periodic boundary in the lattice direction of $\bm{a}_1$ in the pyrochlore lattice to create a pair of open boundary conditions. This operation is equivalent to cut three elastic bonds that are connected to the $C$-site for every supercell.
	}\label{fig6}
\end{figure}

However, cutting a periodic boundary can create a pair of open boundaries, as shown in SI. Fig. \ref{fig6}. This is equivalent to cut three bonds that are connected to the $C$-site for every supercell. The resulting generalized pyrochlore lattice is subjected to open boundary conditions in the primitive direction $\bm{a}_3$ (i.e., boundaries perpendicular to $\bm{b}_3$ are cut open) and periodic boundary conditions in $\bm{a}_1$ and $\bm{a}_2$. The resulting structure can be viewed as a lattice that is periodic in $\bm{a}_1$ and $\bm{a}_2$, with a large unit cell, as shown in Fig. 3\textbf{c} of the main text. Elastic waves in this ``$\bm{a}_1$ and $\bm{a}_2$"-periodic lattice can be classified by surface wavevectors $\bm{k}_{\rm surf} = (\bm{k}\cdot\bm{a}_1, \bm{k}\cdot\bm{a}_2)$ defined in the two-dimensional surface Brillouin zone. Thus, for every large unit cell, the number difference between floppy modes and states of self-stress follow the Calladine-isostatic counting rule $\nu_{\rm fm}-\nu_{\rm sss} = 3$, where $\nu_{\rm fm}$ and $\nu_{\rm sss}$ denote the numbers of floppy modes and states of self-stress per large unit cell (supercell), respectively.

This counting rule can be further extended to the generalized Calladine-isostatic theorem, in which for each surface wavevector $\bm{k}_{\rm surf}$ in the two-dimensional surface Brillouin zone, the number difference between floppy modes and states of self-stress yields 
\begin{eqnarray}\label{countingK}
	\nu_{\rm fm}(\bm{k}_{\rm surf})-\nu_{\rm sss}(\bm{k}_{\rm surf})=3.
\end{eqnarray}
Finally, as there is no states of self-stress in the generalized pyrochlore lattice~\cite{Olaf2016PRL} ($\nu_{\rm sss}(\bm{k}_{\rm surf})=0$), a total of three floppy modes per surface wavevector should emerge from the lattice, with $\nu_{\rm fm}(\bm{k}_{\rm surf})=3$.

Given the surface wavevector $\bm{k}_{\rm surf}$, the three mechanical floppy modes can either localize on the top or bottom boundary of the generalized pyrochlore lattice. Thus, the spatial decay rates along $\bm{a}_3$, namely ${\rm Im\,}(\bm{k}\cdot\bm{a}_3)$ for the three floppy modes, can take different values for given surface wavevector $\bm{k}_{\rm surf}$. Moreover, their signs indicate the localization behaviors of the boundary floppy modes. When ${\rm Im\,}(\bm{k}\cdot\bm{a}_3)<0$, the floppy mode is exponentially localized on the ``top" boundary that $\bm{a}_3$ points to (the normal outward direction of the top boundary is $+\bm{b}_3$), whereas floppy modes with ${\rm Im\,}(\bm{k}\cdot\bm{a}_1)>0$ are localized on the bottom surface (the normal outward direction of the bottom surface is $-\bm{b}_3$). 
The numbers of top and bottom-boundary floppy modes, $\nu_{\uparrow}(\bm{k}_{\rm surf})$ and $\nu_{\downarrow}(\bm{k}_{\rm surf})$, directly correspond to the negative and positive ${\rm Im\,}(\bm{k}\cdot\bm{a}_3)$ solutions in $\det\textbf{C}(\bm{k})=0$, and they yield $\nu_{\uparrow}(\bm{k}_{\rm surf})+\nu_{\downarrow}(\bm{k}_{\rm surf})=\nu_{\rm fm}(\bm{k}_{\rm surf})=3$, in accordance with SI. Eq. (\ref{countingK}).

Thus, the determinant equation, $\det\textbf{C}(\bm{k})=0$, must have three complex-valued $\bm{k}\cdot\bm{a}_3$ solutions for each $\bm{k}_{\rm surf}\in[-\pi,\pi]\times [-\pi,\pi]$. This can be doubly verified by simply computing the determinant of the $12\times 12$ compatibility matrix $\textbf{C}(\bm{k})$ in SI. Eq. (\ref{CM}), which can be symbolically expressed as the third-order polynomial of $e^{\mathrm{i}\bm{k}\cdot\bm{a}_3}$: $\det\textbf{C}(\bm{k})=\alpha e^{-\mathrm{i}\bm{k}\cdot\bm{a}_3}\prod_{l=1}^3 (e^{\mathrm{i}\bm{k}\cdot\bm{a}_3}-\lambda_l)$. Here, $\alpha$ is the coefficient of the polynomial, and $\lambda_l$ for $l=1,2,3$ are the three eigenvalues of the compatibility matrix. $\alpha$ and $\lambda_l$ vary as the surface wavevector $\bm{k}_{\rm surf}$ moves around in the two-dimensional surface Brillouin zone. 
As a result, along the primitive vector direction $\bm{a}_3$, the spatial decay rate of the floppy mode is ${\rm Im\,}(\bm{k}\cdot\bm{a}_3) = -\ln |\lambda_l|$. The number of negative (positive) decay rate solutions, i.e., number of $-\ln|\lambda_l|<0$ ($-\ln|\lambda_l|>0$) solutions, corresponds to the number of floppy modes that are exponentially localized at the top (bottom) boundary of the lattice.

Next, we build the relationship between the signs of $-\ln|\lambda_l|$ and the integer values of the topological winding numbers. To this end, we express the winding numbers as $\mathcal{N}_3(\bm{k}) = 1 - \frac{1}{2\pi\mathrm{i}} \sum_{l=1}^3 \int_{0}^{2\pi} dq_3 \, \partial_{q_3} \ln(e^{\mathrm{i}(\bm{k}\cdot\bm{a}_3+q_3)}-\lambda_l)$. When $|\lambda_l|>1$, the integration contribution is zero to the winding number, whereas for $|\lambda_l|<1$, the contribution is $-1$. Consequently, $\mathcal{N}_3(\bm{k})$ takes the values $1,0,-1$, or $-2$, if three, two, one and zero floppy modes are localized on the bottom boundary of the lattice (zero, one, two, or three floppy modes localizing on the top). At this point, we establish a rigorous mapping between the integer values of the topological winding number and the number of exponentially localized floppy modes on the top and bottom boundaries, which is referred to as the topological bulk-boundary correspondence.

Mechanically isostatic lattices exhibit different topological phases. For instance, the generalized pyrochlore lattice is in the topologically polarized phase, as described in SI. Sec. \uppercase\expandafter{\romannumeral2}(A), and it does not have any gapless Weyl lines. In this phase, the winding numbers remain invariant as $\bm{k}$ traverses the three-dimensional Brillouin zone. Consequently, the localization of boundary floppy modes does not change when the surface wavevector $\bm{k}_{\rm surf}$ is varied in the two-dimensional surface Brillouin zone. In the Weyl phase (SI. Sec. \uppercase\expandafter{\romannumeral2}(B)), isostatic lattices contain gapless lines in the three-dimensional Brillouin zone, where $\det\textbf{C}(\bm{k}_w)=0$, and the floppy modes have a spatial decay rate that approaches zero, ${\rm Im}\,(\bm{k}_w\cdot\bm{a}_3)=0$, at Weyl wavevectors. This means that the wavevector of Weyl floppy modes extends deeply into the lattice. When moving away from the Weyl lines, $\det\textbf{C}(\bm{k})\neq 0$ for real-valued wavevectors in the three-dimensional reciprocal space. In other words, for real-valued surface wavevectors $\bm{k}_{\rm surf} = (\bm{k}\cdot \bm{a}_1, \bm{k}\cdot\bm{a}_2)$ that are not on the surface projection of the Weyl lines, the solution for $\bm{k}\cdot\bm{a}_3$ must be complex. The imaginary part of ${\rm Im\,}(\bm{k}\cdot\bm{a}_3)$ indicates the spatial decay rate of the boundary floppy mode. Since ${\rm Im\,}(\bm{k}\cdot\bm{a}_3)$ becomes zero when $\bm{k}_{\rm surf}$ lies on top of the Weyl line projections, it changes sign as $\bm{k}_{\rm surf}$ moves from the outside to the inside of the Weyl line projections. This indicates a shift in the localization boundary of the corresponding floppy modes, and the jump in the integer value of the winding number $\mathcal{N}_3(\bm{k})$, as $\bm{k}_{\rm surf}$ varies. This indicates that the topological charge of mechanical Weyl lines can be computed by the difference between the winding numbers when $\bm{k}_{\rm surf}$ moves from the outside to the inside of the Weyl line projections. 

\begin{figure}[htbp]
	\includegraphics[scale=0.4]{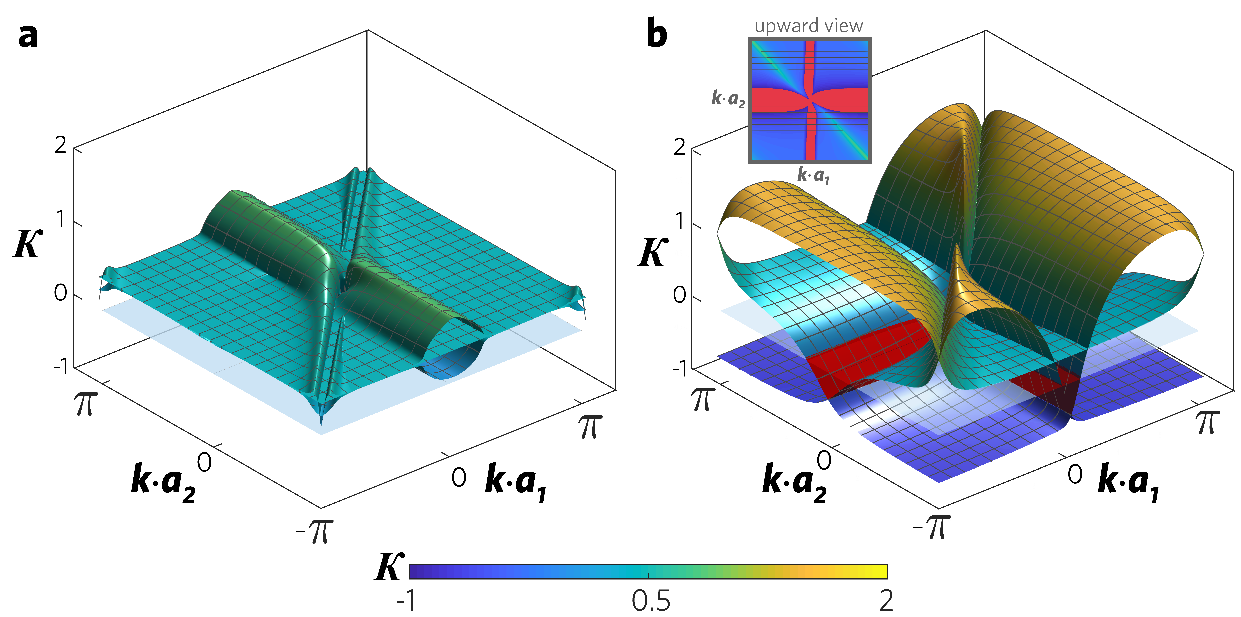}
	\caption{The spatial decay rate of the surface mechanical floppy modes, $\kappa={\rm Im\,}(\bm{k}\cdot\bm{a}_3)$, is plotted in terms of the surface wave numbers $\bm{k}\cdot\bm{a}_1$ and $\bm{k}\cdot\bm{a}_2$. As there are three boundary floppy modes per supercell, three different decay rates are obtained for every surface wavevector $\bm{k}_{\rm surf} = (\bm{k}\cdot\bm{a}_1,\bm{k}\cdot\bm{a}_2)$. (a) The spatial decay rates of the three floppy modes in the topologically polarized phase (unit cell configuration in the inset of Fig. 2\textbf{a}). All three sets of decay rates are consistently positive throughout the entire surface Brillouin zone. 
		(b) The spatial decay rates of the three floppy modes in the mechanical Weyl phase that hosts four Weyl lines (unit cell configuration in the inset of Fig. 2\textbf{e}). The decay rates change from positive to negative as the surface wavevector passes through the surface projections of Weyl lines, as indicated by the inset here. It is notable that the vanishing decay rates constitute the Weyl line projections.}\label{SIfig30}
\end{figure}

In SI. Fig. \ref{SIfig30}, we plot the spatial decay rates of the floppy modes in terms of the surface wavevectors. In the partially-polarized and fully-polarized topological phases, the signs of ${\rm Im\,}(\bm{k}\cdot\bm{a}_3)$ remain invariant in the two-dimensional surface Brillouin zone. In particular, in the fully-polarized topological phase, all decay rates are consistently positive (negative) throughout the entire surface Brillouin zone, indicating the bottom (top) boundary of the pyrochlore lattice that is significantly softer than the opposite parallel surface. This is pictorially manifested in SI. Fig. \ref{SIfig30}\textbf{a}, which corresponds to the geometric parameters shown in SI. Eq. (\ref{deform1}), and the unit cell configuration in Fig. 2\textbf{a} of the main text. All decay rates are positive for the entire 2D surface Brillouin zone, and all floppy modes are localized on the bottom surface. In the four-Weyl-line phase, which corresponds to the geometric parameters in SI. Eqs. (\ref{deform3}) and the unit cell configuration in Fig. 2\textbf{e} of the main text, the decay rate changes from positive to negative as the surface wavevector passes through the surface projection of Weyl lines, and the critical borderline, ${\rm Im\,}(\bm{k}\cdot\bm{a}_3)=0$, corresponds to the surface Brillouin-zone-projections of Weyl lines. This result can be seen in SI. Fig. \ref{SIfig30}\textbf{b}.

Thus, Weyl lines not only affect the topological winding numbers and phases, but also significantly impact the boundary mechanics of isostatic lattices. 
Weyl line projections mark the critical surface wavevectors when the decay rates become zero. The decay rates change signs as the surface wavevector moves from the exterior to the interior of the surface projections of Weyl lines.
This indicates a flip of localized floppy mode from one open boundary to the opposite. Consequently, for mechanically isostatic lattices in the Weyl phases, topologically localized floppy modes arise on a pair of parallel open surfaces, both of which manifest floppy mode-induced softness.


\subsection{Topological phase diagram of the generalized pyrochlore lattice}

\begin{figure}[htb]
	\includegraphics[scale=0.28]{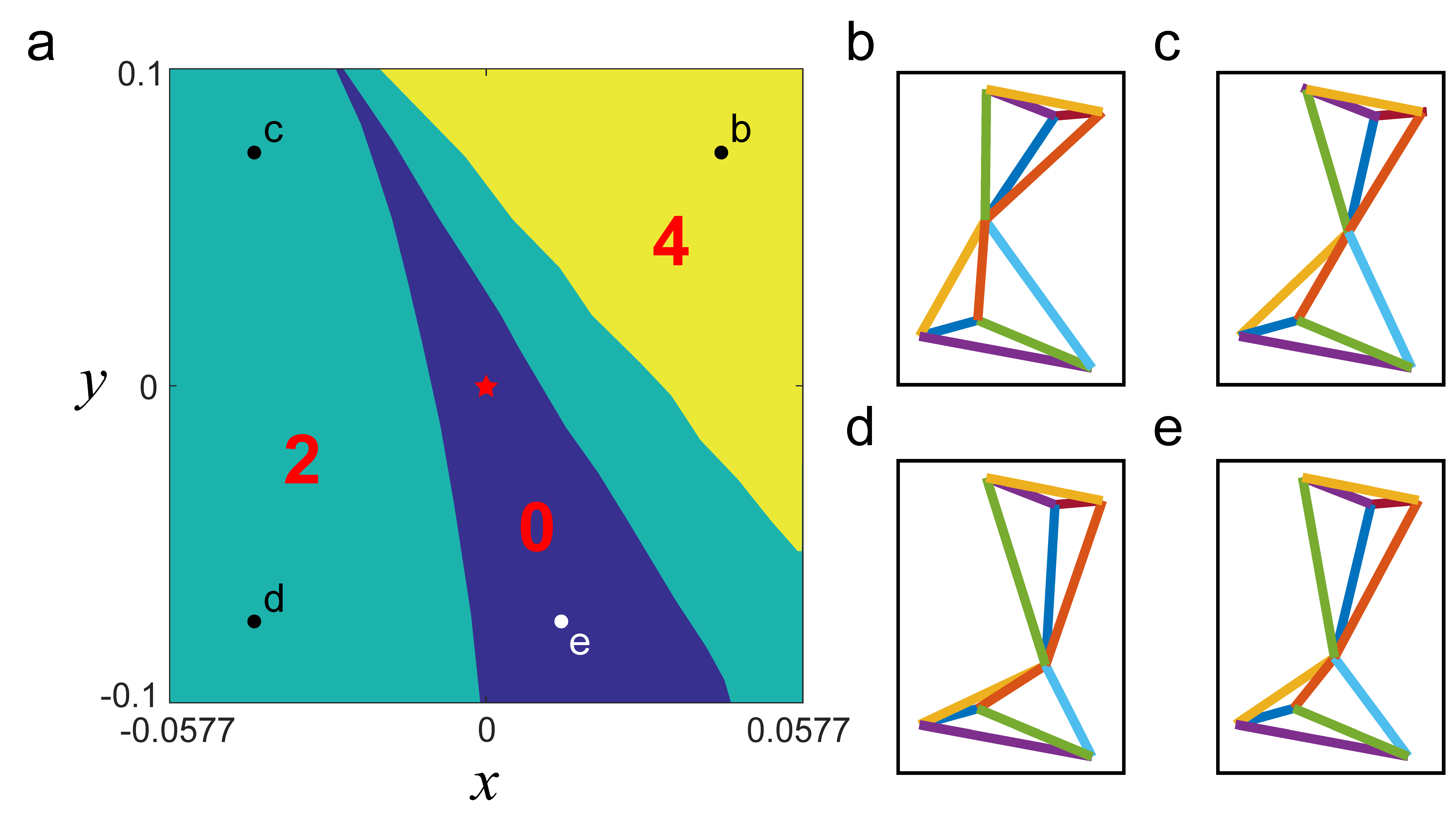}
	\caption{Topological mechanical phase diagram of the generalized pyrochlore lattices as the position of site-C in the unit cell varies. \textbf{a} Yellow, green and blue areas correspond to the geometric configurations of the generalized pyrochlore lattices whose reciprocal spaces have four, two and zero Weyl lines. The asterisk marks the initial configuration of the pyrochlore lattice that corresponds to the fully-polarized topological mechanical lattice in the experiment. \textbf{b}, \textbf{c}, \textbf{d}, and \textbf{e} depict the unit cell configurations for the pyrochlore lattices whose reciprocal spaces have four, two, two, and zero Weyl lines, respectively.
	}\label{SIfig14}
\end{figure}

Finally, we discuss the topological phase diagram of the generalized pyrochlore lattice as the geometric parameters vary. As shown by SI. Fig. \ref{SIfig14}, we construct the topological mechanical phase diagram by varying the geometric parameter, which is the site position of the $C$-vertex. The positions of other sites, including $A$, $B$, and $D$, as well as the primitive vectors $\bm{a}_1$, $\bm{a}_2$, and $\bm{a}_3$, remain the same as those depicted in the unit cell of the topologically fully-polarized configuration.

Specifically, we define the modified $C$-vertex site position, denoted as $C_{\rm phase}$, using the following expression:
\begin{eqnarray}
	C_{\rm phase} = C+x(\bm{a}_1+\bm{a}_2)+y\bm{a}_3.
\end{eqnarray}
Here, $C$ represents the site position of the $C$-vertex in the fully-polarized topological mechanical configuration used in our experiment. $\bm{a}_1=\ell(1,1,0)$, $\bm{a}_2=\ell(0,1,1)$, and $\bm{a}_3=\ell(1,0,1)$ are the primitive vectors of the fully-polarized topological mechanical lattice (as defined in the main text). $x$ and $y$ correspond to the horizontal and vertical axes of the phase diagram, respectively.

The color-coded regions in SI. Fig. \ref{SIfig14} correspond to distinct topological phases, where yellow, green and blue areas indicate the geometric configurations of pyrochlore lattices containing four, two, and zero mechanical Weyl lines in their reciprocal spaces.

We obtain this topological mechanical phase diagram by computing winding numbers (as defined in Eq. (1) of the main text). These winding numbers describe the topological nature of the mechanical band structure and are guaranteed to be integer values. Notably, they exhibit abrupt jumps from one integer to another during topological phase transitions, resulting in the sharp phase boundaries observed.

\section{Mechanical transfer matrix: fully polarize the topological phase of isostatic lattices}

Among various topological phases in mechanically isostatic lattices, including the fully-polarized, partially-polarized, and Weyl phases, the fully-polarized topological phase distinguishes itself from other two, as the boundary mechanical responses are strongly asymmetric in this phase. Thus, it is crucial to design the lattice geometry such that the generalized pyrochlore lattice lies in the fully-polarized topological phase. However, the parameter space of the compatibility matrix, as shown in SI. Eq. (\ref{CM}), is 14-dimensional. It is almost impossible to perform an ergodic scan of all these parameters and pick out the fully-polarized topological phase among all possible phases.

To overcome this challenge, we establish the three-dimensional mechanical transfer matrix that carries out the general design rule of spring-mass isostatic lattices. This analytic transfer matrix polarizes all floppy mode localization to the desired lattice direction by propagating floppy modes through nodes (mass particles) in the mechanical network. Using this methodology, the geometry of the entire structure is polarized to the fully polarized topological phase, and the corresponding floppy modes are consistently driven towards the top or the bottom boundary of the lattice. This analytic transfer matrix significantly reduces the number of adjustable parameters from 14 to 3, making it possible to numerically search the fully-polarized topological mechanical phase in the generalized pyrochlore lattice. 

\begin{figure}[htb]
	\includegraphics[scale=0.4]{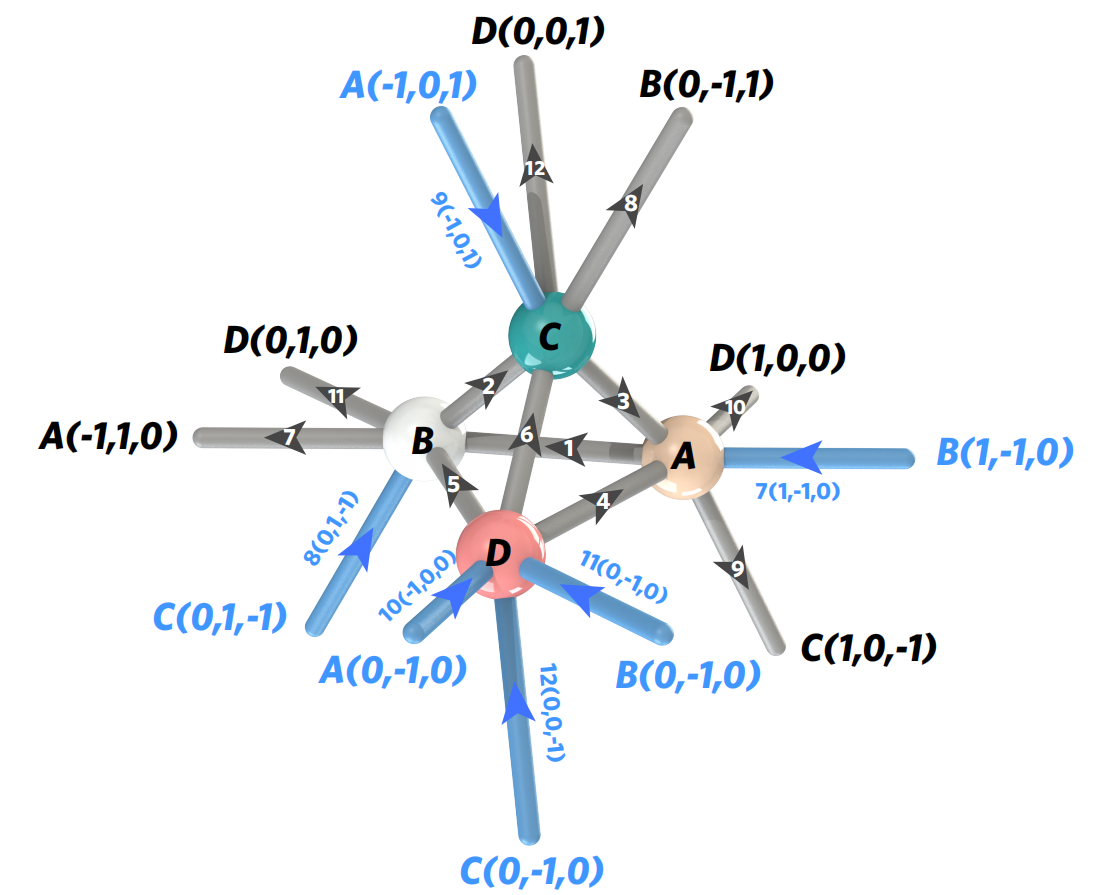}
	\caption{The unit cell of the generalized pyrochlore lattice that contains four sites, marked by $A$, $B$, $C$, and $D$, and 12 bonds, labelled from 1 to 12. The black arrows in the bonds indicate the edge directions. The nearest-neighbor sites are represented by $A(-1,1,0)$, $A(0,-1,0)$, $A(-1,0,1)$, $B(0,-1,0)$, $B(0,-1,1)$, $B(1,-1,0)$, $C(0,1,-1)$, $C(0,-1,0)$, $C(1,0,-1)$, $D(0,1,0)$, $D(0,0,1)$, and $D(1,0,0)$. The nearest-neighbor bonds are marked by $7(1,-1,0)$, $8(0,1,-1)$, $9(-1,0,1)$, $10(-1,0,0)$, $11(0,-1,0)$, and $12(0,0,-1)$ (namely $\bm{l}_7(1,-1,0)$, $\bm{l}_8(0,1,-1)$, $\bm{l}_9(-1,0,1)$, $\bm{l}_{10}(-1,0,0)$, $\bm{l}_{11}(0,-1,0)$, and $\bm{l}_{12}(0,0,-1)$).
	}\label{SIfig10}
\end{figure}

To apply this transfer matrix, we consider the unit cell of the pyrochlore lattice, shown in SI. Fig. \ref{SIfig10}, which contains four nodes (mass particles marked by $A(\bm{n}), B(\bm{n}), C(\bm{n}), D(\bm{n})$) and $12$ (intracell) $+$ $6$ (intercell) Hookean springs. Among these 18 Hookean bonds presented in SI. Fig. \ref{SIfig10}, 12 of them belong to the unit cell and are denoted as $\bm{l}_i(\bm{n})$ for $i=1,2,\ldots, 12$, whereas other 6 bonds come from the elastic springs in the nearest-neighbor unit cells. These bonds are represented by $\bm{l}_7(\bm{n}+(1,-1,0))$, $\bm{l}_8(\bm{n}+(0,1,-1))$, $\bm{l}_9(\bm{n}+(-1,0,1))$, $\bm{l}_{10}(\bm{n}+(-1,0,0))$, $\bm{l}_{11}(\bm{n}+(0,-1,0))$, and $\bm{l}_{12}(\bm{n}+(0,0,-1))$. These 18 elastic springs are connected to the four nodes within the unit cell.

We further analyze the unit cell by decomposing it into four nodes, each of which is linked to six Hookean springs. To be specific, node $A(\bm{n})$ is connected to $\bm{l}_1(\bm{n})$, $\bm{l}_3(\bm{n})$, $\bm{l}_4(\bm{n})$, $\bm{l}_7(\bm{n}+(1,-1,0))$, $\bm{l}_9(\bm{n})$, and $\bm{l}_{10}(\bm{n})$. Node $B(\bm{n})$ is connected to $\bm{l}_1(\bm{n})$, $\bm{l}_2(\bm{n})$, $\bm{l}_5(\bm{n})$, $\bm{l}_7(\bm{n})$, $\bm{l}_8(\bm{n}+(0,1,-1))$, and $\bm{l}_{11}(\bm{n})$. Node $C(\bm{n})$ connects to $\bm{l}_2(\bm{n})$, $\bm{l}_3(\bm{n})$, $\bm{l}_6(\bm{n})$, $\bm{l}_8(\bm{n})$, $\bm{l}_9(\bm{n}+(-1,0,1))$, and $\bm{l}_{12}(\bm{n})$. Node $D(\bm{n})$ is linked to $\bm{l}_4(\bm{n})$, $\bm{l}_5(\bm{n})$, $\bm{l}_6(\bm{n})$, $\bm{l}_{10}(\bm{n}+(-1,0,0))$, $\bm{l}_{11}(\bm{n}+(0,-1,))$, and $\bm{l}_{12}(\bm{n}+(0,0,-1))$. Therefore, each site in the unit cell can be viewed as one node connected to six nearest-neighbor bonds. Below, we establish the analytic mechanical transfer matrix for such ``one node and six bond" system.

\begin{figure}[htb]
	\includegraphics[scale=0.6]{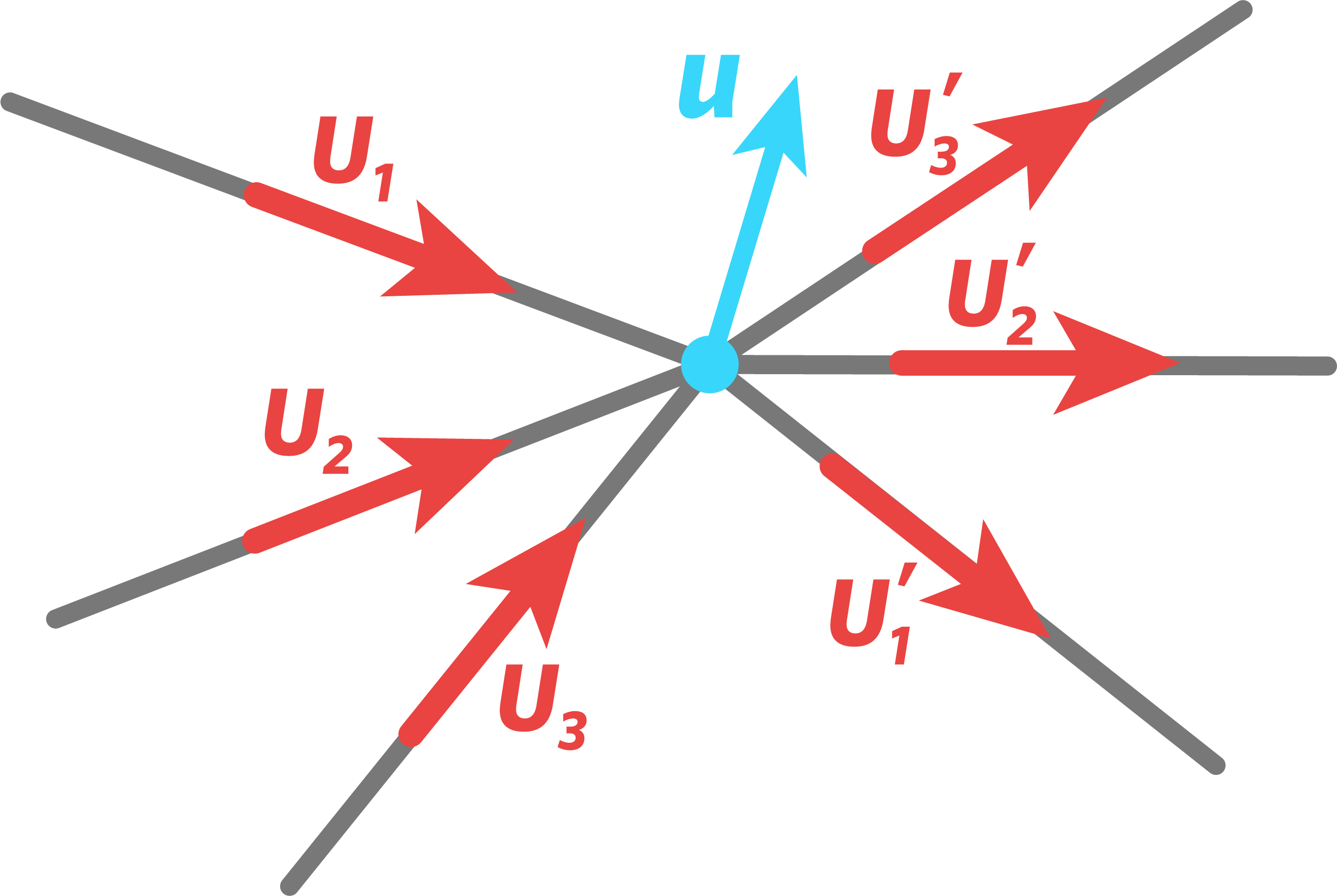}
	\caption{Illustration of the transfer matrix applying on a crosslink. A site that is connected to six bonds, whose site displacement is $\bm{u}$. The bond orientations are $\hat{n}_i$ and $\hat{n}_i'$ for $i=1,2,3$. The longitudinal projections of the displacement vector on the bonds are $U_i = \bm{u}\cdot\hat{n}_i$ and $U_i'=\bm{u}\cdot\hat{n}_i'$ for $i=1,2,3$.
	}\label{SIfig20}
\end{figure}

The mechanical property of the nodes and bonds can be modeled using the three-dimensional mechanical transfer matrix, which is constructed as follows. As shown in SI. Fig. \ref{SIfig20}, a node is connected to six central-force springs. One of the ends of every spring is connected to the node (mass particle), and the bonds are labeled by $i$ or $i'$ for $i=1,2,3$. The unit vectors of the bond orientations are denoted as $\hat{m}_i = (m_{ix}, m_{iy}, m_{iz})$ and $\hat{m}_i' = (m_{ix}', m_{iy}', m_{iz}')$ for $i=1,2,3$. Consider a displacement $\bm{u} = (u_x, u_y, u_z)$ of this mass particle. The site displacement induces elongations on the six bonds, given by $U_i = \bm{u}\cdot \hat{m}_i$ and $U_i' = \bm{u}\cdot \hat{m}_i'$ for $i=1,2,3$. Since we are considering floppy modes, each bond can neither be elongated or abbreviated. Thus, the other end of bond $i$ or $i'$ has to induce a displacement whose longitudinal projection must compensate the elongation caused by $U_i$ or $U_i'$.
As a result, every mass particle serves to ``propagate" the longitudinal projection of node displacements from bond to bond. To compute how floppy mode projections propagate in a network, we establish the relationship between $U_i$ and $U_i'$ by a $3\times 3$ transfer matrix $\mathbf{T}(\hat{m}_1', \hat{m}_2', \hat{m}_3'|\hat{m}_1, \hat{m}_2, \hat{m}_3)$, 
\begin{eqnarray}
	U_i' = \sum_{j=1}^3 \mathbf{T}_{ij}(\hat{m}_1', \hat{m}_2', \hat{m}_3'|\hat{m}_1, \hat{m}_2, \hat{m}_3)U_j
\end{eqnarray}
where every matrix element is given by 
\begin{eqnarray}\label{transfer}
	& {} & \textbf{T}_{ij}(\hat{m}_1', \hat{m}_2', \hat{m}_3'|\hat{m}_1, \hat{m}_2, \hat{m}_3) 
	= \nonumber \\
	& {} & \delta_{ij}+
	\sum_{a,b=1}^3\frac{\epsilon_{jab}}{2}\frac{(\hat{m}_i'-\hat{m}_i)\cdot(\hat{m}_a\times\hat{m}_b)}{\hat{m}_1\cdot(\hat{m}_2\times\hat{m}_3)}
\end{eqnarray}
for $i,j=1,2,3$. In the context of floppy mode considerations, the longitudinal projections of site displacements from bond to bond are related by this transfer matrix. In other words, the nodes that connect to elastic bonds can be viewed as a ``pipeline" that propagates ``the flow of the longitudinal component of floppy mode displacements". Given an arbitrary crystalline or amorphous isostatic network, transfer matrix always solves the floppy mode by computing the longitudinal projections of site displacements throughout the entire network.

Using the analytic technique of transfer matrix, we demonstrate that all floppy modes in the regular pyrochlore lattice are bulk modes. Subsequently, we show that these bulk floppy modes transition to edge modes when the lattice geometry is deformed into the generalized pyrochlore lattice.

In the regular pyrochlore lattice, all filaments (Hookean bonds that connect nearest neighbor sites) form \emph{straight} lines. In this idealized spring-and-mass model, the regular pyrochlore lattice displays an interesting property: all floppy modes are bulk modes. This can be seen in the following analysis. Because all bonds form straight lines, the bond orientations yield the relationship $\hat{m}_i=\hat{m}_i'$ for all nodes. As a result, the transfer matrices shown in SI. Eq. (\ref{transfer}), are $3\times 3$ identity matrices for all nodes throughout the entire network of the regular pyrochlore lattice. Thus, the longitudinal projections of floppy modes yield $U_i'=U_i$ for $i=1,2,3$, which are conserved along each straight line. As the site displacements are the linear superposition of the longitudinal projections of floppy modes, the site movements neither decrease nor increase throughout the entire lattice. In summary, all floppy modes are bulk states in the regular pyrochlore lattice. This is similar to the regular kagome lattice, which has straight filaments that also carry bulk floppy modes~\cite{zhou2018PRL,Zhou2019PRX}. The regular pyrochlore lattice can be seen as a three-dimensional analog of the regular kagome lattice. 

In the two-dimensional generalized kagome lattice, the mechanical frame exhibits static mechanical phases with different topologies where the floppy modes localize at different edges. Can the pyrochlore lattice also exhibit such topological phases? The answer is yes. To this end, we slightly deform the geometry of the regular pyrochlore lattice by bending the straight lines composed of Hookean springs into zigzagged ones. This can be realized by simply changing the orientations of nearest-neighbor bonds to $\hat{m}_i\neq \hat{m}_i'$. Moreover, we ask that the orientation difference, $\delta\hat{m}_i = \hat{m}_i'-\hat{m}_i$, to satisfy the ``slight bending condition" $|\delta\hat{m}_i|\ll 1$, which allows for perturbative expansions on the matrix elements of the mechanical transfer matrix. To the leading order, the transfer matrix can be approximated as 
\begin{eqnarray}
	& {} & \textbf{T}_{ij}(\hat{m}_1', \hat{m}_2', \hat{m}_3'|\hat{m}_1, \hat{m}_2, \hat{m}_3)  \approx \nonumber \\
	& {} & \delta_{ij}\left[1+
	\sum_{a,b=1}^3\frac{\epsilon_{jab}}{2}\frac{\delta\hat{m}_i\cdot(\hat{m}_a\times\hat{m}_b)}{\hat{m}_1\cdot(\hat{m}_2\times\hat{m}_3)}\right].
\end{eqnarray}
Below, we use this first-order approximation to discuss how floppy mode projections evolve from bond to bond in the generalized pyrochlore lattice.

We consider the four sites in the unit cell of the generalized pyrochlore lattice, as shown in SI. Fig. \ref{SIfig10}, in which the unit vectors of the edges are marked as $\hat{t}_i$ for $i=1,2,\ldots,12$. First, consider a site displacement $\bm{u}_D(\bm{n})$, whose longitudinal projections on edges $\bm{l}_{10}(\bm{n}-(1,0,0))$, $\bm{l}_{11}(\bm{n}-(0,1,0))$, $\bm{l}_{12}(\bm{n}-(0,0,1))$, $\bm{l}_4(\bm{n})$, $\bm{l}_5(\bm{n})$, and $\bm{l}_6(\bm{n})$, are $U_{10}(\bm{n}-(1,0,0))$, $U_{11}(\bm{n}-(0,1,0))$, $U_{12}(\bm{n}-(0,0,1))$, $U_4(\bm{n})$, $U_5(\bm{n})$, and $U_6(\bm{n})$. The transfer matrix at site $D(\bm{n})$ relates these displacement projections via
\begin{eqnarray}\label{transD}
	& {} & \left(\begin{array}{ccc}
		U_4(\bm{n})\\
		U_5(\bm{n})\\
		U_6(\bm{n})\\
	\end{array}\right) = \nonumber \\
	& {} & \mathbf{T}(\hat{t}_{4}, \hat{t}_{5}, \hat{t}_{6}|\hat{t}_{10}, \hat{t}_{11}, \hat{t}_{12})
	\left(\begin{array}{ccc}
		U_{10}(\bm{n}-(1,0,0))\\
		U_{11}(\bm{n}-(0,1,0))\\
		U_{12}(\bm{n}-(0,0,1))\\
	\end{array}\right).\qquad
\end{eqnarray}
Given the ``input flow" of the floppy mode projections $U_{10}(\bm{n}-(1,0,0))$, $U_{11}(\bm{n}-(0,1,0))$, and $U_{12}(\bm{n}-(0,0,1))$, transfer matrix solves the ``output flow" of floppy mode projections $U_4(\bm{n})$, $U_5(\bm{n})$, and $U_6(\bm{n})$. Second, the $A(\bm{n})$ site connects six bonds labeled $\bm{l}_4(\bm{n})$, $\bm{l}_7(\bm{n}+(1,-1,0))$, $\bm{l}_9(\bm{n})$, $\bm{l}_{10}(\bm{n})$, $\bm{l}_1(\bm{n})$, and $\bm{l}_3(\bm{n})$. The floppy mode displacement of this $A(\bm{n})$ site can be projected onto these bonds, and these longitudinal projections are related by a transfer matrix
\begin{eqnarray}\label{transA}
	& {} & \left(\begin{array}{ccc}
		U_{10}(\bm{n})\\
		U_1(\bm{n})\\
		U_3(\bm{n})\\
	\end{array}\right)
	=\nonumber \\
	& {} & 
	\mathbf{T}(\hat{t}_{10}, \hat{t}_{1}, -\hat{t}_{3}|\hat{t}_{4}, \hat{t}_{7}, -\hat{t}_{9}) 
	\left(\begin{array}{ccc}
		U_4(\bm{n})\\
		U_7(\bm{n}+(1,-1,0))\\
		U_9(\bm{n})\\
	\end{array}\right).\qquad
\end{eqnarray}
Given the input flow of floppy mode projections, $U_4(\bm{n})$, $U_7(\bm{n}+(1,-1,0))$, and $U_9(\bm{n})$, transfer matrix solves the output flow of $U_{10}(\bm{n})$, $U_1(\bm{n})$, and $U_3(\bm{n})$. Third, the $C(\bm{n})$ site is connected to six bonds, marked as $\bm{l}_3(\bm{n})$, $\bm{l}_6(\bm{n})$, $\bm{l}_8(\bm{n})$, $\bm{l}_{9}(\bm{n}+(-1,0,1))$, $\bm{l}_{12}(\bm{n})$, and $\bm{l}_2(\bm{n})$. The floppy mode projections are related by 
\begin{eqnarray}\label{transB}
	& {} & \left(\begin{array}{ccc}
		U_{9}(\bm{n}+(-1,0,1))\\
		U_{12}(\bm{n})\\
		U_2(\bm{n})\\
	\end{array}\right)=\nonumber \\
	& {} & 
	\mathbf{T}(-\hat{t}_{9}, \hat{t}_{12}, -\hat{t}_{2}|-\hat{t}_{3}, \hat{t}_{6}, -\hat{t}_{8}) 
	\left(\begin{array}{ccc}
		U_3(\bm{n})\\
		U_6(\bm{n})\\
		U_8(\bm{n})\\
	\end{array}\right).\qquad
\end{eqnarray}
This transfer matrix solves the output flow $U_{9}(\bm{n}+(-1,0,1))$, $U_{12}(\bm{n})$, and $U_2(\bm{n})$ provided that the input flow $U_3(\bm{n})$, $U_6(\bm{n})$, and $U_8(\bm{n})$ are given. Finally, the $B(\bm{n})$ site is linked to $\bm{l}_1(\bm{n})$, $\bm{l}_2(\bm{n})$, $\bm{l}_5(\bm{n})$, $\bm{l}_{7}(\bm{n})$, $\bm{l}_{8}(\bm{n}+(0,1,-1))$, and $\bm{l}_{11}(\bm{n})$. The floppy mode projections are related by 
\begin{eqnarray}\label{transC}
	& {} & \left(\begin{array}{ccc}
		U_{7}(\bm{n})\\
		U_{8}(\bm{n}+(0,1,-1))\\
		U_{11}(\bm{n})\\
	\end{array}\right)=\nonumber \\
	& {} & 
	\mathbf{T}(\hat{t}_{7}, -\hat{t}_{8}, \hat{t}_{11}|\hat{t}_{1}, -\hat{t}_{2}, \hat{t}_{5}) 
	\left(\begin{array}{ccc}
		U_1(\bm{n})\\
		U_2(\bm{n})\\
		U_5(\bm{n})\\
	\end{array}\right).\qquad
\end{eqnarray}
Summarizing all these transfer matrices, namely SI. Eqs. (\ref{transD}--\ref{transC}), we use the input flow of floppy mode projections $U_7(\bm{n}+(1,-1,0))$, $U_8(\bm{n})$, $U_9(\bm{n})$, $U_{10}(\bm{n}-(1,0,0))$, $U_{11}(\bm{n}-(0,1,0))$, $U_{12}(\bm{n}-(0,0,1))$ to obtain the output flow of floppy mode projections $U_7(\bm{n})$, $U_8(\bm{n}+(0,1,-1))$, $U_9(\bm{n}+(-1,0,1))$, $U_{10}(\bm{n})$, $U_{11}(\bm{n})$, $U_{12}(\bm{n})$.

We now incorporate the geometric deformations, namely $\Delta A = (\Delta A_x, \Delta A_y, \Delta A_z)$, $\Delta B = (\Delta B_x, \Delta B_y, \Delta B_z)$, $\Delta C = (\Delta C_x, \Delta C_y, \Delta C_z)$, and $\Delta D = (\Delta D_x, \Delta D_y, \Delta D_z)$, to the initial geometry of the regular pyrochlore lattice. Therefore, the bond vectors in the pyrochlore lattice are changed from $\bm{l}_i$ to $\bm{l}_i+\Delta\bm{l}_i$. For example, the bond vector $\bm{l}_1$ is now changed into $\bm{l}_1+\Delta B-\Delta A$, as indicated by SI. Eqs. (\ref{1}). This allows us to reach a new ground state called the generalized pyrochlore lattice. As the geometric changes are small comparing to the bond lengths, the ``bending angles", i.e., orientation difference between nearest neighbor bonds, are very small. Using the first-order approximation in $\Delta\bm{l}_i$, we have 
\begin{eqnarray}\label{changeL}
	\delta \hat{t}_i = \hat{t}_{i+6}-\hat{t}_i \approx -2\left(\frac{\Delta \bm{l}_i}{|\bm{l}_i|}-\frac{(\bm{l}_i\cdot\Delta\bm{l}_i)\bm{l}_i}{|\bm{l}_i|^3}\right)
\end{eqnarray}
for $i = 1,2,\ldots, 6$, where $\bm{l}_i$ is the bond vector of the $i$th edge, and $\Delta\bm{l}_i$ is the change in the bond vector between the generalized and regular pyrochlore lattices. Substituting SI. Eq. (\ref{changeL}) into SI. Eqs. (\ref{transD}--\ref{transC}) significantly simplifies the recursion relations among the longitudinal projections of site displacements:
\begin{eqnarray}
	& {} & (1-\sqrt{2}\delta t_{1y})U_7(\bm{n}+(1,-1,0)) = U_7(\bm{n}),\nonumber \\
	& {} & (1-\sqrt{2}\delta t_{2z})U_8(\bm{n}+(0,1,-1)) = U_8(\bm{n}),\nonumber \\
	& {} & (1-\sqrt{2}\delta t_{3x})U_9(\bm{n}+(-1,0,1)) = U_9(\bm{n}),\nonumber \\
	& {} & (1+\sqrt{2}\delta t_{4z})U_{10}(\bm{n}-(1,0,0)) = U_{10}(\bm{n}),\nonumber \\
	& {} & (1+\sqrt{2}\delta t_{5x})U_{11}(\bm{n}-(0,1,0)) = U_{11}(\bm{n}),\nonumber \\
	& {} & (1+\sqrt{2}\delta t_{6y})U_{12}(\bm{n}-(0,0,1)) = U_{12}(\bm{n}).
\end{eqnarray}

Now, we ask that the floppy mode projections consistently and exponentially decrease from the bottom boundary (i.e., the lattice boundary whose normal is $-\bm{b}_3$) to the top boundary (i.e., the surface normal is $+\bm{b}_3$). To be specific, we demand $U_8(\bm{n})<U_8(\bm{n}+(0,1,-1))$, $U_9(\bm{n}+(-1,0,1))<U_9(\bm{n})$, and $U_{12}(\bm{n}+(-1,0,1))<U_{12}(\bm{n})$, leading to the constraints
\begin{eqnarray}\label{constraint}
	\Delta A_{x}>\Delta C_{x}, \quad \Delta C_{y}>\Delta D_{y},\quad {\rm and}\quad \Delta B_{z}>\Delta C_{z}.\quad
\end{eqnarray}
Under these constraints, all floppy modes tend to exponentially localize on the bottom boundary of the pyrochlore lattice. SI. Eq. (\ref{constraint}) significantly reduces the 14 adjustable parameters in the compatibility matrix to only 3. The geometric parameters used in the fully-polarized topological phase in SI. Sec. \uppercase\expandafter{\romannumeral2}(C), directly follow the constraints provided by SI. Eqs. (\ref{constraint}), with the other parameters, including $\Delta A_{y}$, $\Delta A_{z}$, $\Delta B_{x}$, $\Delta B_{y}$, $\Delta D_{x}$, and $\Delta D_{z}$, are randomly generated by our numerical simulations.

Our work realizes, for the first time, the topologically fully polarized isostatic lattices in three dimensions (3D). This achievement enables novel scientific and application advancements not possible for one- and two-dimensional isostatic metamaterials, as detailed below in SI. Sec. \uppercase\expandafter{\romannumeral5} and \uppercase\expandafter{\romannumeral6}, respectively.


\begin{figure}[htbp]
	\includegraphics[scale=0.45]{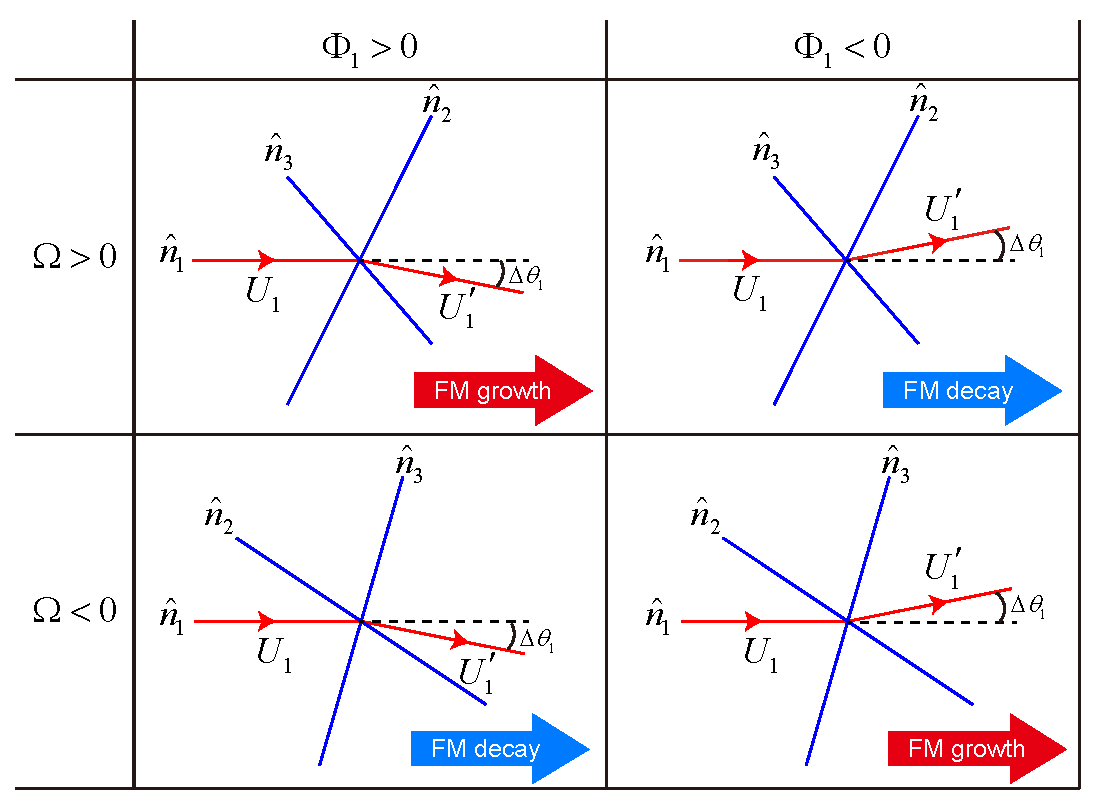}
	\caption{The spatial evolution of mechanical floppy modes (abbreviated as FM) for different geometric configurations of the junctions that are connected to six bonds. 
	}\label{figR13}
\end{figure}

Here, we give a brief argument to intuitively understand how $\Delta A$, $\Delta B$, $\Delta C$, and $\Delta D$, as well as $\Delta\bm{a}_{i=1,2,3}$ are chosen to realize the full polarization of mechanical topology. In SI. Fig. \ref{figR13}, we exemplify one such node that is connected to six springs. We can think of this ``one-node, six-spring" junction as a ``one-node, three-filament" system, because we design the red bonds such that the bending angle is small comparing to the crossing angles between the filaments. We consider the spatial evolution of the floppy mode from bond to bond in the central red filament, which is mathematically derived by the mechanical transfer matrix. SI. Fig. \ref{figR13} itemizes the spatial growing and decaying behaviors of the floppy mode in the central red filament when it passes through the junction. The spatial growing or decaying behavior is governed by the sign of $\Phi_1 = \sum_{a,b=1}^3 \epsilon_{iab}\delta\bm{m}_i\cdot(\bm{m}_a\times \bm{m}_b)/2$, namely the bending angle of the central fiber, and the sign of $\Omega = \bm{m}_1\cdot(\bm{m}_2\times\bm{m}_3)$, namely the crossing angles between neighboring filaments.

Based on the design principle presented in SI. Fig. \ref{figR13} for every single junction, we ask that in the pyrochlore lattice, floppy modes consistently grow from the top layer to the bottom layer when they pass through all junctions. This is realized by designing the sign of bending angles of the filaments for every junction, which in turn is controlled by site positions and primitive vectors such that every node in the pyrochlore isostatic lattice consistently increases the floppy mode when it propagates from bond to bond in the downward vertical direction. Fig. 1 of the main text illustrates one such geometric configuration. Interestingly, we find that, by changing the site positions only, while the primitive vectors stay unchanged, the isostatic lattice is already capable of increasing the floppy mode from the top boundary to the bottom boundary.

In conclusion, SI. Fig. \ref{figR13} presents the design principle of the full polarization of mechanical topology in isostatic lattices, such as the pyrochlore lattice. By following this rule, we modify site positions, namely $\Delta X$ with $X=A,B,C,D$, and primitive vectors, namely $\Delta\bm{a}_i$ with $i=1,2,3$, to ensure that the floppy mode consistently grow from the top surface to the bottom boundary. While Fig. 1 of the main text illustrates a design with the $\Delta\bm{a}_i=0$ condition, this is not a necessary condition for the full polarization of mechanical topology.

\section{Scientific Advancements from the Topological pyrochlore Lattice}

\subsection{Topologically insulated mechanical phase in 3D}

While previous studies~\cite{PhysRevLett.122.248002, paulose2015PNAS} have managed to create 3D isostatic structures, their systems always exhibit mechanical bulk and surface conduction. Our pyrochlore lattice is the first isostatic structure that demonstrates the fully-gapped and fully-polarized topological mechanical phase in 3D. This results in the mechanical insulation for both the \emph{bulk and top-surface} of the pyrochlore lattice.

\begin{figure}[htbp]
	\includegraphics[scale=0.28]{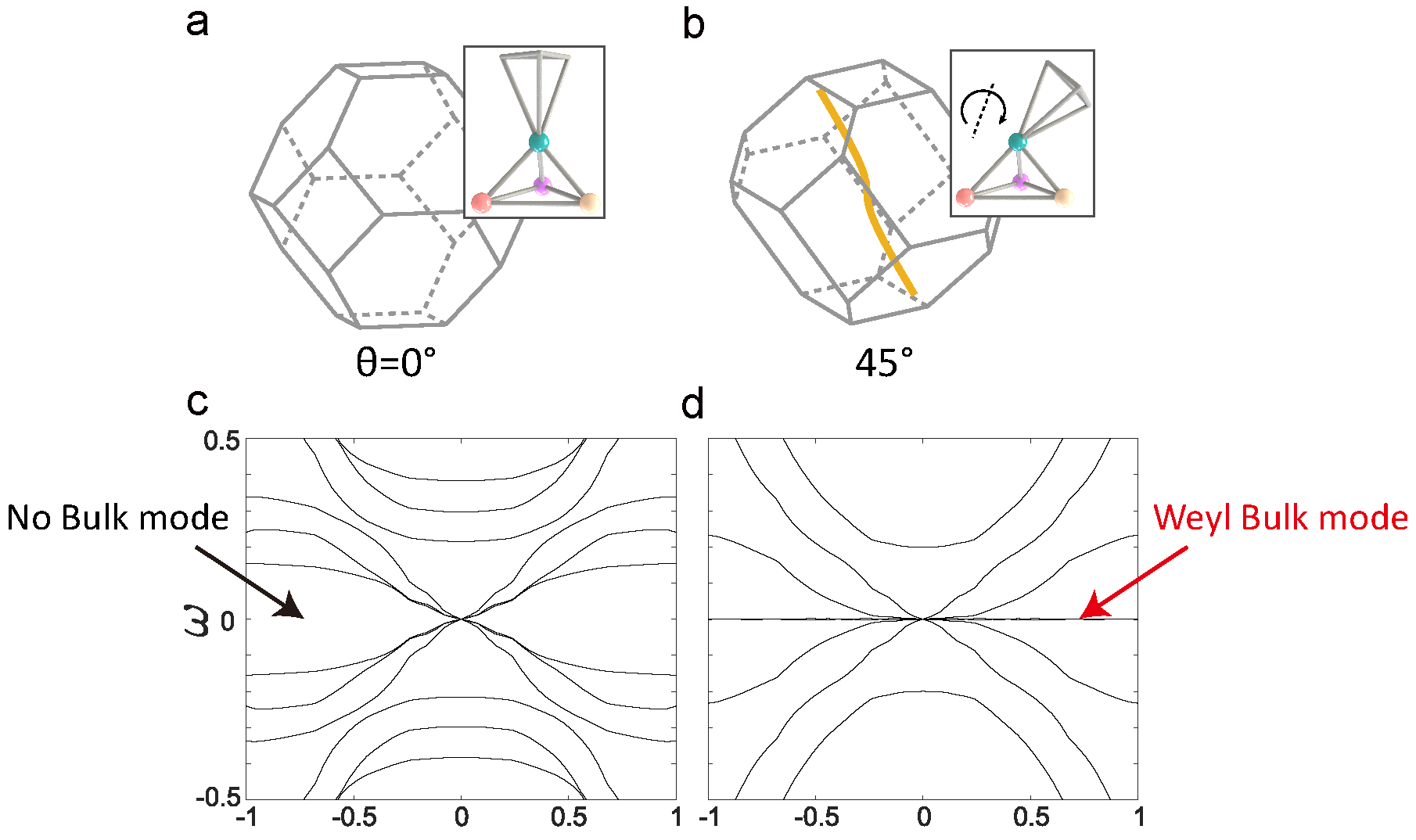}
	\caption{\textbf{a} and \textbf{b}, Brillouin zones of the pyrochlore lattice in the topologically-polarized and Weyl phases, respectively. \textbf{c} and \textbf{d} are the mechanical band structures for \textbf{a} and \textbf{b}, respectively.}\label{figR1}
\end{figure}

To elucidate the fully-gapped nature of the generalized pyrochlore lattice, we establish the Schr\"{o}dinger-like mechanical Hamiltonian by taking the ``square root" of the Newtonian dynamical matrix~\cite{kane2014topological}. This mechanical Hamiltonian reads
\begin{eqnarray}\label{R1}
	\textbf{H} = \left(\begin{array}{cc}
		0 & \textbf{C}\\
		\textbf{C}^\top & 0\\
	\end{array}\right),
\end{eqnarray}
where $\textbf{C}$ is referred to as the compatibility matrix. We numerically compute the eigenvalues in Eq. (\ref{R1}) using Bloch boundary conditions. We then plot the mechanical band structures for the topologically polarized and Weyl phases in Figs. \ref{figR1}\textbf{c} and \textbf{d}, respectively. In both cases, we compute the band structures for wavevectors that follow the orange mechanical Weyl line defined in SI. Fig. \ref{figR1}\textbf{b}. 

The zero-frequency Weyl modes shown in SI. Fig. \ref{figR1}\textbf{d} differentiate the band structures between SI. Figs. \ref{figR1}\textbf{c} and \textbf{d}, with SI. Fig. \ref{figR1}\textbf{c} representing bulk mechanical insulation for the topologically polarized phase and SI. Fig. \ref{figR1}\textbf{d} displaying bulk mechanical conduction for the Weyl phase. Furthermore, the results in Figs. 3 and 4 in the main text demonstrate that in the fully-polarized topological phase, the top boundary of the pyrochlore lattice is entirely free of floppy modes, leading to \emph{top surface} mechanical insulation. As a result, mechanical conduction is solely due to the \emph{bottom boundary} of the topological pyrochlore lattice. These unique features of the fully-polarized topological phase hold promise for a range of future applications.

For instance, the mechanical conduction that occurs solely at one boundary of the fully-polarized topological lattice may transform into \emph{chiral} floppy modes when the intrinsic spins of phonon waves are considered. This fundamental mechanism, namely the mechanical spin-momentum locking~\cite{long2018intrinsic}, has recently been experimentally realized in 2D isostatic lattices~\cite{cheng2023backscattering} that exhibits backscattering-free acoustic propagation. \emph{Chiral} floppy modes, if realized in 3D systems, could revolutionize the field of isostatic metamaterials by enabling low-frequency sound filtering, novel transducer designs, and mechanical signal processing modulators. Moreover, these modes could facilitate the classification of strong and weak mechanical topological insulators, which are novel and unique concepts in 3D systems. These ideas draw inspiration from similar classifications in 3D electronic topological insulators~\cite{PhysRevLett.98.106803}.

\subsection{Surface topological floppy modes robust against disorders}

Topological materials should exhibit robustness against disorder, allowing their unique edge modes to persist even when periodicity is disrupted. Here we show robust surface (2D) topological floppy modes against disorders in 3D systems. To incorporate spatial disorder in the generalized pyrochlore lattice, we shift their site positions to new coordinates using the formula
\begin{eqnarray}\label{R2}
	X(n_1,n_2,n_3) = X+[1+R\,\, {\rm rand}(n_1,n_2,n_3)]\delta X.
\end{eqnarray}
Here, $X$ represents the four sites $A, B, C, D$ in the unit cell, $(n_1,n_2,n_3)$ denote the three lattice indices that mark the position of the unit cell. $\delta X$ corresponds to the geometric change of the regular pyrochlore lattice that leads to the topologically polarized phase (see their definitions in the main text). ${\rm rand}(n_1,n_2,n_3)$ is the random number ranging from $-1$ to $1$, and $R$ represents the disorder strength. As the disorder strength rises, the compatibility matrix and the mechanical Hamiltonian (see SI. Eq. (\ref{R1})) change accordingly.

\begin{figure}[htbp]
	\includegraphics[scale=0.36]{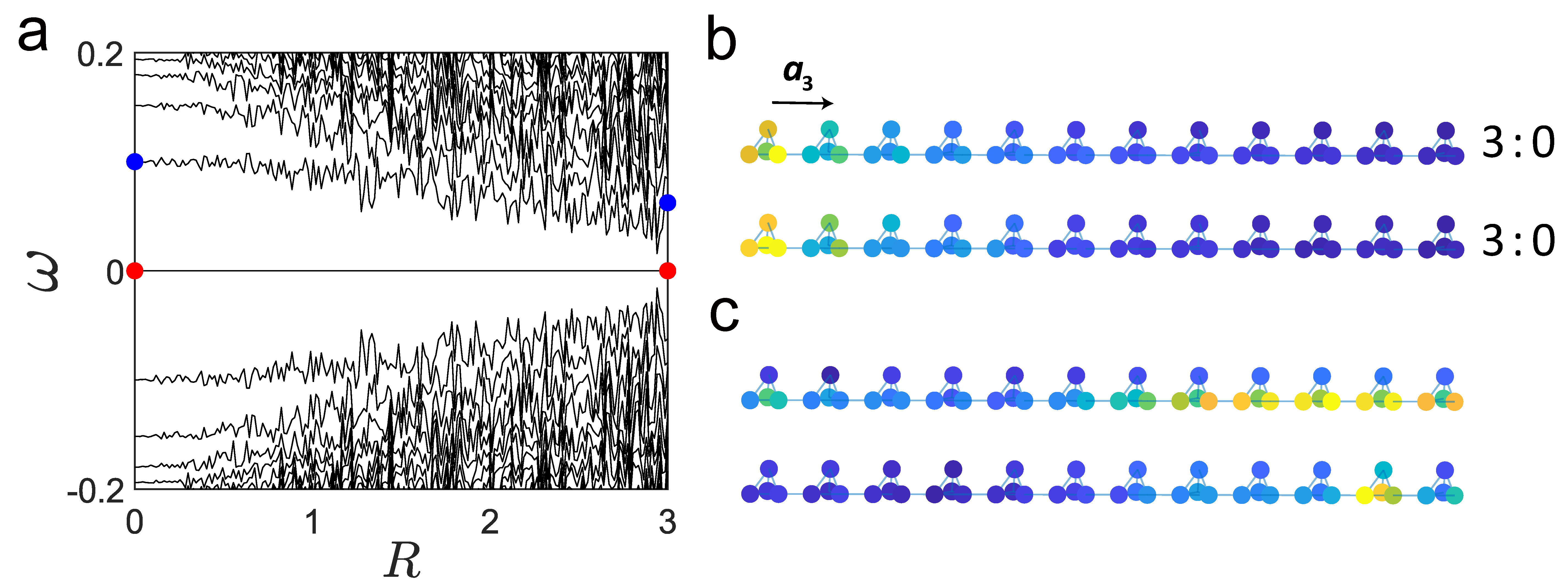}
	\caption{\textbf{a}: Mechanical band structure of the topologically-polarized supercell pyrochlore lattice as the disorder strength increases. \textbf{b}: Spatial profiles of the summation of floppy mode weight for zero disorder in the top panel and strong disorder in the bottom panel. \textbf{c}: Spatial profiles of the bulk phonon modes for zero disorder and strong disorder in the top and bottom panels, respectively. The phonon frequencies are marked by the blue dots in \textbf{a}.
	}\label{figR2}
\end{figure}

We calculate the eigenvalues of the mechanical Hamiltonian with increased disorder strength to investigate its impact on the band structure. A surface wavevector of $\bm{k}=\bm{b}_1/10+\bm{b}_2/4$ is chosen to illustrate the band structures in SI. Figs. \ref{figR2}\textbf{a} for the topologically polarized pyrochlore lattice. SI. Figs. \ref{figR2}\textbf{b} depicts the summation of floppy mode weight in the supercell pyrochlore lattice, with the number of floppy modes localized on the top and bottom open surfaces marked beside these plots. 

Floppy modes exhibit topological robustness. This is evidenced by both the opening of the band gap in SI. Fig. \ref{figR2}\textbf{a}, and the spatial profile of floppy mode weight in SI. Fig. \ref{figR2}\textbf{b} that remains insensitive to an increase in disorder strength ($\nu_\downarrow:\nu_\uparrow=3:0$ remains unchanged as disorder increases, where $\nu_\downarrow$ and $\nu_\uparrow$ denote the number of floppy modes localized on the bottom and top boundaries).

The 3D fully-polarized topological lattice is fundamentally distinct from its 2D counterparts, such as the kagome lattice. As we highlight in the following discussions, the out-of-plane displacements in 2D lattices lack topological robustness~\cite{charara2022omnimodal, zunker2021soft}, but they manifest topological polarization and protection in our 3D lattice.

In summary, our design exhibits unprecedented mechanical properties in the 2D surface of 3D pyrochlore lattice, which is stable against disturbances to the internal structure of the lattice.

\subsection{Asymmetric propagation of elastic waves}

\begin{figure}[htbp]
	\includegraphics[scale=0.4]{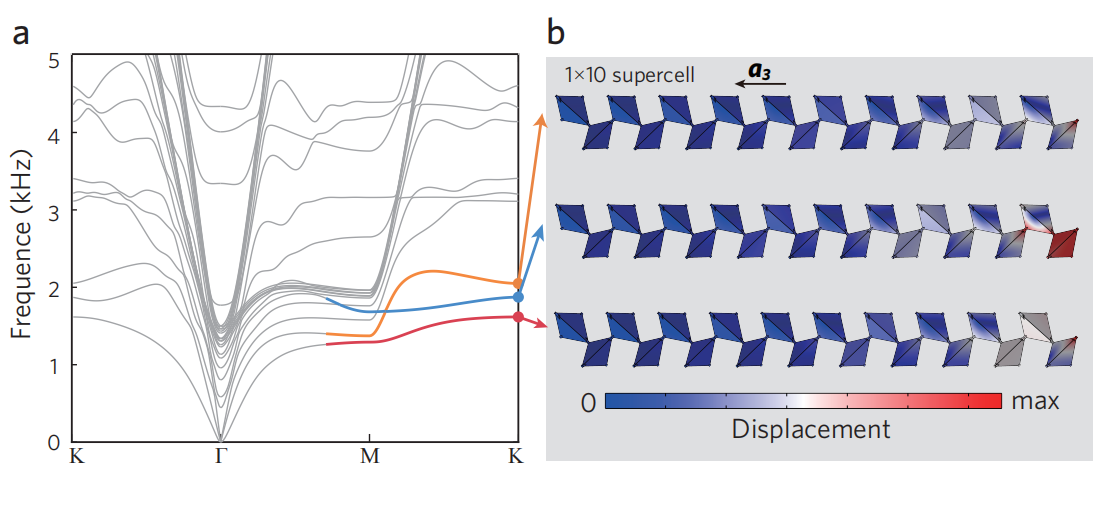}
	\caption{COMSOL simulations for asymmetric acoustic wave propagation in the topologically polarized pyrochlore lattice, where the lattice constant $\ell=20mm$, and the hinge diameter is $1mm$. \textbf{a}: Band structure of a supercell lattice that is composed of 10 unit cells. The supercell lattice is subjected to open boundary conditions in the $\bm{a}_3$-direction, and Bloch boundary conditions in $\bm{a}_1$ and $\bm{a}_2$. \textbf{b}: Spatial profiles of the displacement fields for the lowest three acoustic modes, where the surface wavevector is illustrated at the high-symmetry point $\bm{k}=K=(5\bm{b}_1+\bm{b}_2)/6$.
	}\label{figR3}
\end{figure}

In the topologically fully-polarized phase, all mechanical floppy modes exclusively localize on the bottom boundary of the lattice. This directly leads to the mechanism called \emph{mechanical amplification}, as floppy modes exponentially grow from the rigid top boundary to the soft bottom boundary. More importantly, when finite bending stiffness is considered, this amplification mechanism further extends the asymmetric mechanical propagation to non-zero frequencies. This asymmetric acoustic propagation for in-plane motions has been studied in 2D topologically polarized lattices~\cite{PhysRevLett.121.094301}, but the out-of-plane displacements lack such asymmetry~\cite{charara2022omnimodal, zunker2021soft}.

In our 3D lattice, asymmetric acoustic wave propagation is enabled in all three spatial dimensions. This is numerically demonstrated in SI. Fig. \ref{figR3}\textbf{a}, where the small but finite bending stiffness of COMSOL-constructed topological pyrochlore lattice raises the frequencies of ``floppy modes" to finite values (and therefore termed as ``soft modes" now), introducing a non-zero group velocity for these topological soft modes. Due to the mechanism of mechanical amplification, soft modes propagate in an asymmetric way between the top and bottom surfaces, as shown by SI. Fig. \ref{figR3}\textbf{b}.

\subsection{Static mechanical non-reciprocity in fully-polarized isostatic lattices}

Reciprocity is a fundamental principle for linear physical systems that obey time-reversal symmetry. It guarantees that the transmissibility function between any two points in space remains identical, irrespective of material or geometric asymmetries. This principle also remains valid for our fully-polarized topological pyrochlore lattice in the linear elastic regime.

By breaking the symmetry in the transmissibility, an enhanced control over signal transport, isolation, and vibration protection can be attained. To achieve non-reciprocal transmission of physical quantities, one needs to either break time-reversal symmetry or include nonlinear effects~\cite{zhou2022NC, zhou2021JMPS}. So far, most devices break reciprocity using dynamical systems that break time-reversal symmetry. Here, we show that our fully-polarized metamaterial can break static reciprocity when nonlinear elasticity (large geometric deformation) is considered.

\begin{figure}[htbp]
	\includegraphics[scale=0.4]{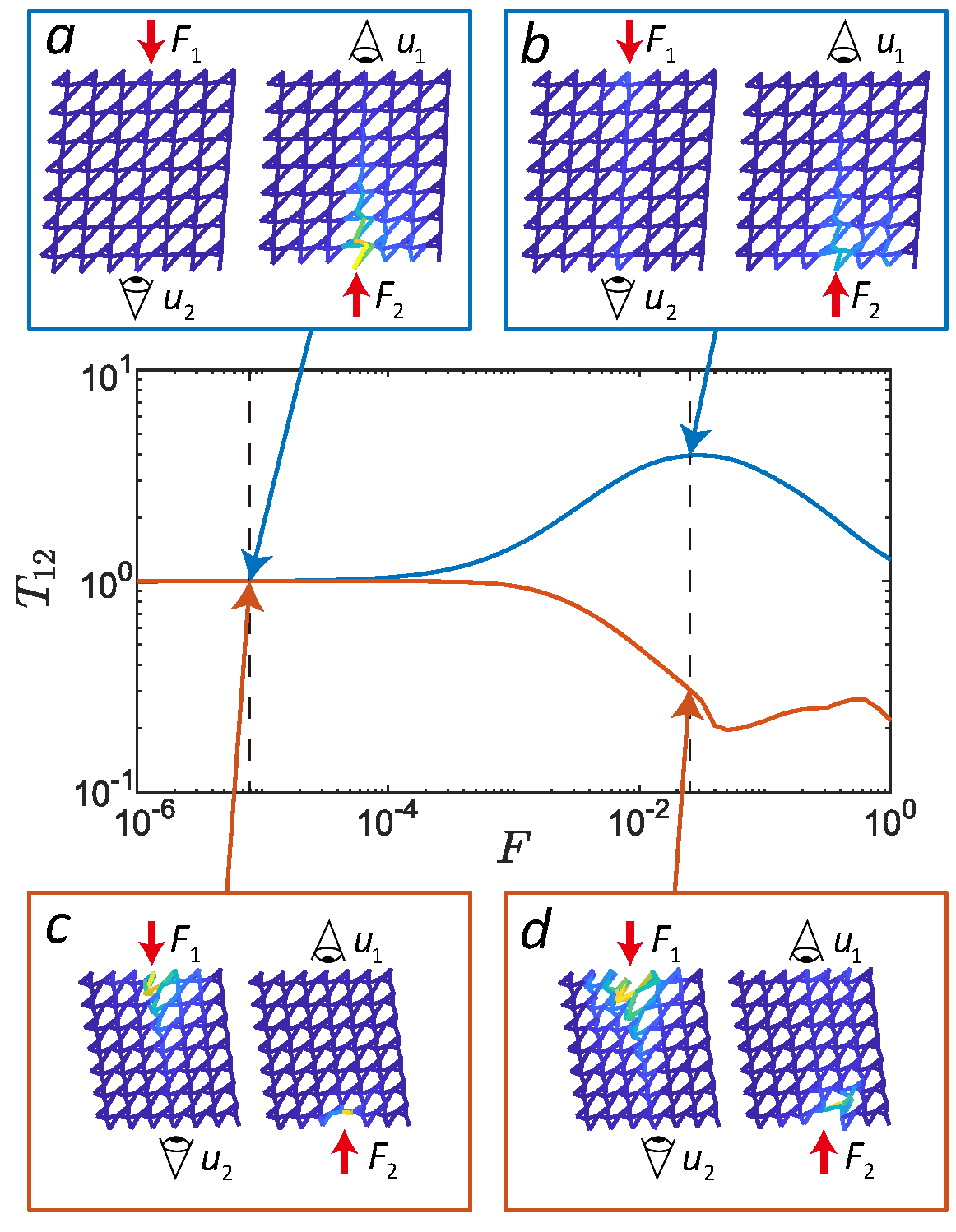}
	\caption{Static mechanical non-reciprocity in the 3D isostatic lattice, where the blue and red curves describe the mechanical transmissibility ratio in the topologically polarized and Weyl phases, respectively. \textbf{a} and \textbf{c}, The resulting lattice deformations in the topologically polarized and Weyl phases, respectively, where the small external force is $F=8\times 10^{-6}$. The site displacements are enlarged by a factor of 1000 for visual convenience. \textbf{b} and \textbf{d}, The lattice deformations for large external force $F=2.5\times 10^{-2}$. The site displacements are enlarged by $3$ times.}\label{figR4}
\end{figure}

To this end, we numerically construct a lattice consisting of $7\times 7 \times 8$ unit cells, with open boundaries on the top and bottom, and fixed boundaries on the four side surfaces. The lattice constant, Hookean spring strength, and particle mass are all set to unity. Every particle is subjected to a weak on-site potential with the pinning spring constant $k_0=10^{-3}$. The fully-polarized topological mechanical lattice (SI. Fig. \ref{figR4}\textbf{a}) can be transformed by the Guest-Hutchinson mode with an angle of $45^\circ$ to reach the Weyl phase (SI. Fig. \ref{figR4}\textbf{c}). Below, we numerically address (linear mechanical) reciprocity and (nonlinear mechanical) non-reciprocity in both the topologically polarized and Weyl phases.

To study mechanical non-reciprocity, we introduce ``mechanical transmissibility" from point 1 to 2 in space, denoted by 
\begin{eqnarray}\label{R3}
	\chi_{12} = u_2/F_1.
\end{eqnarray}
The choice of points 1 and 2 is arbitrary, but we exemplify them as the $C$-tips of the central unit cells on the top and bottom open surfaces. As shown in SI. Fig. \ref{figR4}\textbf{a}, $F_1$ is the input static force at point 1, and is in the normal direction of the top surface. $u_2$ is the normal projection of the resulting displacement at point 2. Similarly, we define $\chi_{21} = u_1/F_2$ as the mechanical transmissibility from 2 to 1. Finally, we define the ``ratio of mechanical transmissibility",
\begin{eqnarray}\label{R4}
	T_{12} = \chi_{12}/\chi_{21},
\end{eqnarray}
that quantifies the the level of mechanical non-reciprocity.

Linear elastic systems obey reciprocity by showing $T_{12}=1$. This stems from the \emph{time-reversal symmetric} Green’s function of linear response theory. We numerically confirm this mechanical reciprocity ($T_{12}=1$) in the small-force regime ($10^{-7}\le F\le 10^{-5}$) using the algorithm of molecular dynamics. As the resulting site displacements are much less than the lattice constant (see SI. Fig. \ref{figR4}\textbf{a}), this force range defines the ``linear elastic regime", where both the topologically polarized and Weyl phases exhibit mechanical reciprocity.

However, nonlinear mechanics showcases strong non-reciprocity. Here, nonlinear effects only stem from geometric nonlinearity, because Hookean springs do not have material nonlinearity. This ``nonlinear elastic regime" is reflected by SI. Figs. \ref{figR4}\textbf{b} and \textbf{d} within the regime of large input force $10^{-4}\le F\le 1$, in which the resulting lattice deformations are comparable to the lattice constant. In this regime, $T_{12}$ (the ratio of mechanical transmissibility) significantly deviates from 1, reflecting the strongly asymmetric mechanical transmissibility function.

In the topologically fully-polarized phase, $T_{12}\gg1$ (blue curve of SI. Fig. \ref{figR4}), indicating that the mechanical transmissibility from the topologically rigid top boundary to the soft bottom boundary is drastically higher than in the reversed direction. This non-reciprocity stems from the nonlinear stiffening effect of the soft boundary, because this stiffening effect is much less significant for the rigid boundary. In the Weyl phase, $T_{12}\ll1$, reflecting that the mechanical transmissibility from the top to bottom is much less than that in the reversed direction (orange curve in SI. Fig. \ref{figR4}). This is because, in the Weyl phase, the stiffening effect has a larger impact on the soft top surface than the bottom one.

In summary, when fully-polarized topological stiffness meets nonlinear effects, our 3D isostatic lattice serves as the ideal device for static non-reciprocal mechanics. Moreover, the mechanical transmissibility significantly drops from $T_{12}\gg 1$ to $T_{12}\ll 1$ when the lattice is reversibly transformed from the topological phase to the Weyl phase by the Guest-Hutchinson mode.

When bending stiffness is considered, floppy modes evolve into topological soft modes with non-zero frequencies. As a result, our topological pyrochlore lattice can exhibit dynamical non-reciprocity for all three spatial dimensions, in contrast to the 2D topological isostatic lattices that manifest dynamical non-reciprocity \emph{only for in-plane motions}.

\subsection{Other future scientific directions}

\emph{Higher-order topological mechanical floppy modes---} Recent developments in electronic topological insulators have unveiled a new class of topological states, namely higher-order topological modes, that localize not on the edges of the system but on its corners and hinges~\cite{benalcazar2017quantized}. However, in 2D isostatic lattices, the design of higher-order topological mechanics requires a gapped mechanical band structure at the $\Gamma$-point, which necessitates breaking translational symmetry by placing mass points on fixed and frictionless rails~\cite{sarkar2023mirror}, substantially complicating the lattice structure.

Our 3D design of isostatic metamaterials may resolve this gapless problem in the mechanical band structure, because unlike 2D isostatic lattices, a finite wavenumber $k_z$ in the vertical direction can break translational symmetry in the horizontal directions, and ultimately gaps the mechanical bands at the $\Gamma$-point.

\emph{Topological mechanical zero modes bounded by dislocations---} Topologically protected mechanical floppy modes can be localized around lattice dislocations. This effect stems from the interplay between two Berry phases: the Burgers vector of the dislocation, and the topological polarization of the isostatic lattice. This dislocation-bounded softness has been studied in both 2D and 3D isostatic lattices~\cite{paulose2015NP, Vitelli2017PNAS}. However, to date, the 3D study is only theoretical, as the lattice required to achieve dislocation-bounded floppy modes is a stacked kagome lattice with a highly complex structure that cannot be realized experimentally.

Our pyrochlore lattice is the first 3D topological metamaterial that can be experimentally realized. Based on the dislocation design in Ref.~\cite{Vitelli2017PNAS}, the pyrochlore metamaterial may experimentally realize dislocation-bounded mechanical floppy modes. This approach has broad applications, such as the mechanical information storage and mechanical logic gates.

\emph{Machine Learning meets topological static mechanics---} Building upon our pyrochlore structure, there is a growing need to explore new lattice structures 
which also exhibit fully-polarized topological mechanics. This searching process can be practically implemented through Machine Learning, where we can train the algorithm to learn and classify isostatic lattices based on their topological mechanical phases, which include fully-polarized, partially-polarized, and Weyl phases.

By doing so, we anticipate that novel designs of isostatic lattices with improved mechanical polarization can be achieved. This, in turn, may lead to the creation of topological soft metamaterials that can maintain their unique properties even when larger bending stiffness is present. The development of such metamaterials can significantly enhance their potential applications, as illustrated in SI. Sec. \uppercase\expandafter{\romannumeral6} below.

\section{Potential applications of the topological pyrochlore lattice}

The applications of topological floppy modes have been extensively investigated in 2D isostatic lattices, including their use in preventing fracturing caused by defects and damage, mechanical information storage, and the creation of transformable topological mechanical metamaterials. However, these benefits fail to apply to out-of-plane motions~\cite{charara2022omnimodal, zunker2021soft}. Here, we show that the polarized surface mechanics in 3D pyrochlore lattice enables new applications not possible in 2D lattices.

\subsection{Topologically protected all-terrain tire}

Our 3D-printed topological isostatic metamaterial boasts high versatility and can be applied to various areas. One promising application is in the development of airless and lightweight tires. Fig. 4\textbf{e} shows a topological pyrochlore lattice arranged in a cylindrical domain, creating a highly porous wheel. The system is designed so that the surface with topological softness faces outward, while the boundary with topological rigidity folds inward to securely attach to the axle without any shaking or instability.

Our topological isostatic metamaterial tire can roll smoothly over rough terrain, thanks to its outer surface that contains topological softness. This is demonstrated in Fig. 4\textbf{f}, which displays the preliminary numerical results of the tire's displacement response while rolling on a rugged terrain profile. The distinct advantage of this design is that the inner layers of the metamaterial retain their topological softness, even under severe conditions such as extreme temperatures or friction that can wear out the outer layers. As a result, the tire remains functional even if the outer layers are damaged, setting it apart from traditional (inflatable) tires that become non-functional when punctured by a sharp object.

\subsection{Topologically switchable landing gear for aircrafts}

Thanks to the freely adjustable Guest-Hutchinson modes in isostatic lattices, our 3D pyrochlore metamaterial can be reversibly switched between topologically polarized and Weyl phases. The distinct topological surface elasticity of this mechanical metamaterial makes it an excellent candidate for designing reconfigurable landing gears for drones and other aircrafts.

As depicted in Fig. 4\textbf{g} of the main text, the pyrochlore landing gear can be switched to the topologically polarized configuration when the drone is in the air. In this configuration, the topologically protected states of self-stress make the bottom surface rigid, providing topological rigidity that minimizes swaying. As the drone makes a landing, the pyrochlore landing gear transitions to the Weyl phase. In this phase, the surface softness is localized on the bottom boundary of the lattice, which efficiently absorbs the shock with ground.

Our novel pyrochlore metamaterial design allows for versatile and controllable manipulations of surface elasticity, which offers significant implications for the design and deployment of aircrafts in various environments.

\subsection{Topological elasticity in 3D}

\begin{figure}[htbp]
	\includegraphics[scale=0.8]{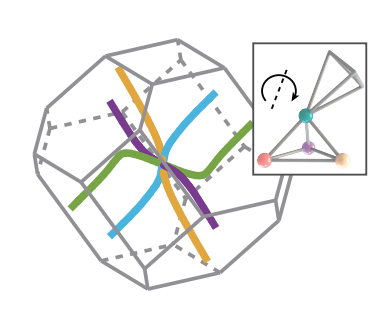}
	\caption{Mechanical Weyl lines pass through the $\Gamma$-point of the Brillouin zone.}
	\label{figR8}
\end{figure}

Topological mechanics has been focused on discrete lattice systems that require the knowledge of their microscopic details. Recent studies~\cite{mao2020PRL} extend topological mechanics to ``topological elasticity", where one- and two-dimensional topological isostatic media can be designed based on only a few macroscopic elastic parameters without knowing their microscopic designs.

However, this theory is not applicable to 3D, because in 3D isostatic media, mechanical Weyl lines can pass through the $\Gamma$-point of the Brillouin zone (as shown in SI. Fig. \ref{figR8}). This results in the presence of long-wavelength mechanical Weyl modes in 3D continuum isostatic media, which have been precluded in 1D and 2D topological elastic theory.

Moreover, in 3D isostatic media, mechanical Weyl modes may induce interesting an-isotropic elastic response, because 3D isostatic lattices have been known to exhibit an-isotropic mechanics that depends on the choice of the surface wavevector. This may intrigue novel designs of isostatic media in 3D, with directional and polarized surface elasticity.

\subsection{Other potential applications of 3D isostatic metamaterials}

\emph{Fracturing protected by topological states of self-stress in isostatic lattices---} It has been demonstrated in Refs.~\cite{paulose2015PNAS, zhang2018NJP} that in 2D isostatic lattices, states of self-stress in domain walls of topologically distinct isostatic lattices can prevent catastrophic fracturing, even with significant damage or small defects. In contrast to brittle materials~\cite{griffith1921vi} where stress usually accumulates at crack tips and causes avalanches, fracturing in topological isostatic lattices with self-stress domain walls occurs gradually, avoiding large avalanches. Thus, isostatic lattices show great potential in shock absorption and structural reinforcement, thanks to their controllable fracturing. However, these benefits were only studied in 2D. Our 3D prototype addresses the limitation in isostatic lattices for protecting materials against damage and fracturing.

\emph{Topological mechanical cloak---} In fully-polarized topological isostatic lattices, site displacements are exponentially reduced from the soft surface to the rigid one, resulting in minimal deformation on the topologically rigid surface. This mechanism enables a novel application known as the topological mechanical cloak, which protects materials from elastic deformation. By utilizing micro-machining techniques~\cite{buckmann2014elasto}, we can fabricate the topological pyrochlore lattice at the nanoscale. This not only enables the creation of a precise mechanical cloaking system but also increases its effectiveness in blocking particle displacements on the soft side.

\emph{Self-adaptive metamaterials for soft robotics---} Soft robotics has led to a range of diverse applications, including soft wearable devices and medical implants~\cite{lin2022soft}. Despite significant progress in this field, several challenges remain, including the connection problem in modular soft robotic systems, structural limitations of the robots themselves, and instability-induced loss of control during high-speed motion.

Using topological isostatic metamaterials with switchable, highly polarized surface mechanics, one can address these issues. Our metamaterial, for instance, offers high strength and controllable fracturing properties, enabling more stable module connections. Additionally, due to its high adaptability and tunability, this material is excellent for supporting all-terrain robotic motions, opening new opportunities for enhancing the performance of soft robotic systems.

\end{appendices}

\end{document}